\documentclass[preprintnumbers]{article}
\usepackage{amssymb,amsmath,graphicx,epsfig,subfig,color,hyperref,url}
\usepackage[margin=1in]{geometry}
\usepackage{tabu}
\usepackage{cite}
\usepackage{caption}
\usepackage{mathptmx}
\usepackage{wrapfig}
\usepackage{epsfig}
\usepackage{footnote}

\def\beq {\begin{equation}}
\def\eeq {\end{equation}}
\def\bea {\begin{eqnarray}}
\def\eea {\end{eqnarray}}
\def \PMET{\rm p{\!\!\!/}_T}

\newcommand{\br}{\begin{eqnarray}}
\newcommand{\er}{\end{eqnarray}}
\newcommand{\be}{\begin{equation}}
\newcommand{\ee}{\end{equation}}

\newcommand{\gsim}{\raisebox{-0.13cm}{~\shortstack{$>$ \\[-0.07cm] $\sim$}}~}

\def \PMET{p{\!\!\!/}_T}
\newcommand{\comment}[1]{}

\usepackage{color}
\usepackage{hyperref}
\usepackage{url} 
\hypersetup{linktocpage=true}
\usepackage[all]{hypcap}
\hypersetup{
   bookmarks=true,         
   unicode=false,          
   pdftoolbar=true,        
   pdfmenubar=true,        
   pdffitwindow=false,     
   pdfstartview={FitH},    
   pdftitle={My title},    
   pdfauthor={Author},     
   pdfsubject={Subject},   
   pdfcreator={Creator},   
   pdfproducer={Producer}, 
   pdfkeywords={keyword1} {key2} {key3}, 
   pdfnewwindow=true,      
   colorlinks=true,       
   linkcolor=blue,          
   citecolor=magenta,        
   filecolor=magenta,      
   urlcolor=cyan           
}
\usepackage{fancyhdr}
\begin{document}
\begin{flushright}
{\bf{ LAPTH-045/13 }}
\end{flushright}

\begin{center}
{\LARGE\bf
Probing the flavor violating scalar top quark signal at the LHC} \\[5mm]
\bigskip
{\large\sf Genevieve Belanger} $^{a},$
{\large\sf Diptimoy Ghosh} $^{b}$, 
{\large\sf Rohini Godbole} $^{c}$, 
{\large\sf Monoranjan Guchait} $^{d}$
and
\newline
{\large\sf Dipan Sengupta} $^{d}$ 
\\ [4mm]
\bigskip
{\noindent $^{a)}$ 
LAPTH, Univ. de Savoie, CNRS,\\
\hspace*{0.1in}  B.P.110, F-74941 Annecy-le-Vieux, France }  \\

{\noindent $^{b)}$ 
INFN, Sezione di Roma,\\
\hspace*{0.1in} Piazzale A.  Moro 2, I-00185 Roma, Italy. }  \\
 
\medskip

{\noindent $^{c)}$ 
Center for High Energy Physics, 
Indian Institute of Science,  \\
\hspace*{0.1in} Bangalore, 560012, India. }

\medskip

{\noindent $^{d)}$ 
Department of High Energy Physics, 
Tata Institute of Fundamental Research,  \\
\hspace*{0.1in} 1, Homi Bhabha Road, Mumbai 400 005, India.  }
\end{center}


\begin{abstract} 
\begin{large}
 The Large Hadron Collider(LHC) has completed its
run at 8 TeV with the experiments
ATLAS and CMS having collected about 25 $\rm fb^{-1}$
of data each.  Discovery of a light Higgs boson, coupled with lack of evidence for supersymmetry at the LHC so far, has motivated studies of supersymmetry in the context of naturalness with the principal focus being the third generation squarks. 
In this work, we analyze the prospects of the 
flavor violating decay mode $\rm \tilde{t}_1\to c\chi_{1}^{0}$
at 8 and 13 TeV center of mass energy at the LHC.
This channel is also relevant in the dark matter context
for the stop-coannihilation scenario,  
where the  relic density depends on the mass difference between 
the  lighter stop quark ($\tilde{t}_1$) and the lightest
neutralino($\chi_{1}^{0}$) states.
This channel is extremely challenging to probe,
specially for situations when the mass difference between
the lighter stop quark  and the lightest
neutralino is small. 
Using certain kinematical properties of signal 
events we find that the level 
of backgrounds can be  reduced substantially. We find that 
 the prospect for this channel is limited 
due to the low  production cross section for top squarks and
limited luminosity at 8 TeV, but at the 13 TeV LHC with
100 $fb^{-1}$ luminosity,  it is possible to probe top 
squarks with masses  up to  $\sim$ 450 GeV. 
We also discuss how the sensitivity could be
significantly improved by tagging charm jets.
\end{large}
\end{abstract}

\newpage

\begin{large}
\section{Introduction}
\label{intro}
At the end of the 8 TeV run the Large Hadron Collider(LHC)
has collected about 25 $\rm fb^{-1}$ of data and produced 
the most defining observation since the onset of its 
operation in 2008. This observation is the finding of a
new boson at 125 GeV\cite{:2012gu,:2012gk}, which in all
its glory seems to behave like the standard model(SM) Higgs boson.
However, its true nature will be 
 revealed only after precision measurements of  
 its couplings and spin as well as their comparison with SM predictions.
Interestingly, this discovery of the Higgs like boson has galvanized the 
beyond standard model(BSM) physics hunters in analyzing  
its  impact on various new physics models including supersymmetry(SUSY).
The implications of the 125 GeV Higgs on SUSY have been well 
studied in the literature 
\cite{Baer:2011ab,Akula:2011aa,Feng:2011aa,
Heinemeyer:2011aa,Buchmueller:2011ab,Draper:2011aa,
Cao:2011sn,Hall:2011aa,Ellis:2012aa,Baer:2012uya,
Maiani:2012ij,Cheng:2012np,Cao:2012fz,
Brummer:2012ns,Balazs:2012qc,Feng:2012jf,
Ghosh:2012dh,Fowlie:2012im,Athron:2012sq,
CahillRowley:2012rv,Akula:2012kk,Cao:2012yn,
Arbey:2012dq,Nath:2012fa,Ellis:2012nv,Chakraborti:2012up,Arbey:2012bp,Chakraborty:2013si,Dighe:2013wfa,Arbey:2013fqa,Belanger:2013pna,Djouadi:2013uqa}, and this is sure  to continue  in the coming days. 

The early searches in SUSY, the most popular model among all the 
BSM  options,  have 
focussed, both phenomenologically 
\cite{Chatterjee:2012qt,Howe:2012xe,Arbey:2012fa,Baer:2012vr}
 and experimentally\cite{CMS-PAS-SUS-12-005,ATLAS-CONF-2012-109},
on the constrained minimal
supersymmetric standard model(CMSSM)/minimal 
supergravity(mSUGRA)\cite{Cremmer1978231,Cremmer1979105,PhysRevLett.49.970,
PhysRevD.27.2359,Nath1983121,Ohta01081983}.
The initial searches probed mainly the gluino
($\tilde{g}$) and the squarks($\tilde{q}$)
of the first two generations. 
The current lower limits on the gluino and the first two generations squark masses
in the framework of CMSSM stand at $\rm m_{\tilde{g}} \gsim 1.5~TeV$ for almost degenerate
gluino and squarks and $\rm m_{\tilde{q}} > 1.4~TeV$ for very high squark masses,
\cite{Aad:2013wta,ATLAS-CONF-2013-047,ATLAS-CONF-2013-061}.

The  LHC constraints on the third generation squarks, stops($\tilde t_{1,2}$) and 
sbottoms($\tilde b_{1,2}$) are  weaker  because of the
lower production cross-section for a squark pair of a single flavor and are weakened even further when the  mass difference between 
the squark and the neutralino is small.  
The negative results in the initial searches for the gluino and first two generations of squarks, together with the discovery of a 'light' Higgs boson have prompted consideration of natural SUSY, 
and hence a light third generation squark
as an attractive framework for phenomenological
studies. 

Natural SUSY \cite{Cohen:1996vb,Hall:2011aa,Feng:2012jf,
Berger:2012ec,Cao:2012rz,Randall:2012dm,Espinosa:2012in,Baer:2012mv}
is inspired by the observation
that the most relevant superparticles  
responsible for cancellation of quadratic divergences in 
the Higgs loop corrections  are stop quarks. 
At the loop level the Higgs potential 
receives corrections from gauge 
and Yukawa interactions, the dominant
part being the top-stop loop. The radiative
corrections ( $\delta m_{H_{u}}^{2}$) are
quadratically proportional to the
third generation left and right 
soft masses along with the trilinear
coupling $\rm A_t$. Thus the trilinear
couplings and the soft masses 
which control the third generation
spectrum determine the level of fine tuning
required to stabilize the Higgs mass 
in the theory.  Another measure of 'naturalness' is 
also inspired by demanding that the level of fine tuning among different terms in the relation connecting $M_{Z}^{2}, \mu^{2}$ 
and  the SUSY  breaking scale is not high. These
considerations thus imply light Higgsinos or light third generation squarks.
The natural SUSY spectrum  thus requires only a few particles 
below the TeV scale,  the stops and the sbottom, the lighter charginos and neutralinos, and the gluino  
\cite{Papucci:2011wy,Espinosa:2012in,Hall:2011aa,Baer:2012mv}.
These natural SUSY scenarios have been 
studied in the literature for a
wide range of phenomenological models 
and signatures. Furthermore, the anatomy 
of lighter third generation superparticles
has been  well dissected in the literature 
in terms of models and collider
signatures 
\cite{Desai:2011th,He:2011tp,Drees:2012dd,Berger:2012ec,Plehn:2012pr,Han:2012fw,Barger:2012hr,
Choudhury:2012kn,Cao:2012rz,Chen:2012uw,Bornhauser:2010mw,Kraml:2005kb,Yu:2012kj,Ajaib:2011hs,Ghosh:2012ud,
Dutta:2012kx,Alves:2012ft,Berenstein:2012fc,Ghosh:2012wb,Chakraborty:2013moa,
Ghosh:2013qga,Han:2013kga,Buchmueller:2013exa,Kribs:2013lua,Kowalska:2013ica}.  

Most of these studies, especially in the context of  colliders 
have focused on the decay of lighter stop($\tilde{t}_1$) and bottom($\tilde{b}_1$), 
$\rm \widetilde
{t}_{1} \to t \chi_{1}^{0} $ and 
$\rm \widetilde{b}_{1} \to b \chi_{1}^{0}$ respectively. It has
been observed that the feasibility of these
channels depend critically on the mass
difference 
\br
\rm \Delta m= m_{\widetilde
{t}_{1}(\widetilde{b}_{1})} - m_{\chi_{1}^{0}},
\label{eq:dm}
\er
since it determines the hardness of the 
final state particles. 
A large value of $\Delta m$ leads to 
large jet momentum and a sizable missing transverse momentum ($\rm \PMET$), both of 
which are imperative to suppress the SM backgrounds.
Interestingly in the large $\Delta m$ scenario,
the use of jet-substructure and top tagging 
is also an important tool to suppress backgrounds
\cite{Kaplan:2012gd,Plehn:2011sj}.
To alleviate the problem of low mass differences,
shape analysis with various kinematic variables have 
also been considered \cite{Alves:2012ft}.

As noted earlier the non-observation 
of a light sparticle in strongly interacting sector, i.e,
a  gluino and/or a squark of first two generation, in the 
sub-TeV regime  along with the observation of a light Higgs  have prompted ATLAS and CMS to perform dedicated searches for the third generations squarks.  Note that, the limits on the 
masses of
the squarks of first two generations using the generic SUSY searches are not applicable to the case of third generation squarks. 
 The  dedicated searches look for third generation squarks produced directly via QCD processes as well as those produced indirectly in gluino decays, in  a plethora of final 
states coming from a variety of decay channels of $\tilde{t_1}$ 
assuming specific mass relationships among $\rm \widetilde{g}$ ,
$\rm \widetilde{t}_1$, $\rm \chi_{1}^{\pm}$ and the
$\rm\chi_{1}^{0}$, assumed to be the lightest SUSY particle(LSP).
The principle decay channels
studied are $\rm \widetilde
{t}_{1} \to t \chi_{1}^{0} $ and  $\rm \widetilde
{t}_{1} \to b \chi_{1}^{\pm} $, with leptons and 
b jets in the final state from top quark decay. ATLAS searched for 
$\tilde{t}_1$ in the decay channel, $\rm 
\tilde{t}_{1} \to t \chi_{1}^{0} $, from direct stop pair
production using 8 TeV LHC data 
with $\rm 13~fb^{-1}$ luminosity and ruled out $m_{\tilde{t}_1}$ 
between 225 GeV- 575 GeV for an LSP mass 
up to 175 GeV \cite{ATLAS-CONF-2012-166}.
ATLAS also probed $\tilde{t}_1$ in the
channel $\rm \widetilde{t}_{1}\to b\chi_{1}^{\pm}\to b\chi_{1}^{0}ff'$    
assuming $\rm \Delta m =m_{\chi_{1}^{\pm}}-m_{\chi_{1}^{0}}$ to be
5 GeV and 20 GeV. For $\rm \Delta m=5~GeV$ they ruled out stop masses    
of about 600 GeV in a corridor of lightest neutralino mass
\cite{ATLAS-CONF-2013-001}.
 With the use of the kinematic variable $\rm M_{T2}$,
ATLAS also excluded $m_{\tilde{t}_1}$ in the range
of 150 GeV-450 GeV for the channel 
$\rm \widetilde
{t}_{1} \to b \chi_{1}^{\pm} $, where the chargino is nearly
degenerate with the $\tilde{t}_1$ state\cite{ATLAS-CONF-2012-167}. 
Similarly, CMS searched for $\tilde{t}_1$ in the channels 
$\rm \widetilde
{t}_{1} \to t \chi_{1}^{0} $ and  $\rm \widetilde
{t}_{1} \to b \chi_{1}^{\pm} $, and excluded it
with a mass between 160 GeV -430 GeV for a LSP 
mass up to 150 GeV \cite{CMS-PAS-SUS-12-023}.

However, when the lighter stop becomes next to lightest SUSY particle(NLSP), the stop searches at colliders are quite different and become challenging. In  this scenario,
the dominant decay modes are via the flavor changing  decays  and the four  body decay \cite{PhysRevD.36.724,Muhlleitner:2011ww,Boehm:1999tr},
\begin{eqnarray}
\tilde{t}_1 & \to & c\chi_{1}^{0}  \\
\label{eq:2jets}
            & \to & bff'\chi_{1}^{0}.
\label{eq:4body}
\end{eqnarray}
Stop pair production followed by these decays leads to final states
containing a heavy quark pair $c \bar c$ or $b \bar b$ respectively 
and are given by:
\begin{equation}
\rm pp \to \widetilde{t_{1}}\widetilde{t_{1}^{*}}\to c\bar{c} +2\chi_{1}^{0}\to 2 jets + \PMET,
\label{eq1}
\end{equation}  
\begin{equation}
pp \to \widetilde{t_{1}}\widetilde{t_{1}^{*}}\to b\bar{b} + 2\chi_{1}^{0} 
+ 2ff'.
\label{eq2}
\end{equation}
The flavor violating decay mode yields precisely two jets and missing transverse
momenta($\PMET$) due to the presence of $\chi_{1}^{0}$. 
The relative decay rates into  the above two channels are extremely 
sensitive to the model parameters \cite{Boehm:1999tr,Djouadi:2001dx}. The 
signal sensitivity for the four body decay channels 
has been studied  for different parameters\cite{Das:2001kd}.

For very low values of $\Delta m$ (Eq.~\ref{eq:dm}) it is rather 
difficult to obtain a reasonable signal sensitivity. In this case, the strategies have been 
to look at the mono-jet + $\PMET$ \cite{He:2011tp,Drees:2012dd}
and mono-photon+$\PMET$ final state \cite{Carena:2008mj,Belanger:2012mk}.
 The experimental limits on these channels come from reinterpretation of the  monojet searches in ATLAS and CMS 
\cite{ATLAS-CONF-2011-096,Chatrchyan:2012me}.
With the available data at 7 TeV, lighter stops of mass below 
about 200 GeV are excluded for the above mentioned decay channels.
\comment{
\textcolor{blue}{Thus it can be observed that the limits on the masses of stops from these
channels are rather limited which can be attributed to
the fact that the final state objects are 
soft due to low value of $\Delta m$ leading to a lower acceptance of 
signal events.
Therefore, it is a challenging task to probe these channels
for very low $\Delta m$ cases.}}
Thus the limits on the mass of the lightest stop from these
channels are rather weak. This is because  the final state objects are 
soft due to the low value of $\Delta m$ leading to a lower acceptance of 
signal events.
Therefore, it is a challenging task to probe these channels
for very low $\Delta m$ cases.

In this work we \comment{\textcolor{blue}{ try to explore}}
explore the possibility to find a
signal for $\tilde{t_1}$ in the flavor violating decay, 
Eq. \ref{eq1}, resulting in a di-jet +$\PMET$
signature.
Note that the flavor violating decay mode 
is also important in the dark matter context
in the stop co-annihilation scenario \cite{Boehm:1999tr}. The 
correct relic density abundance in this case 
crucially depends on $\Delta m$. 
In analyzing this signal we apply different types of kinematic selection
cuts which are described in the following sections. For the case of 
flavour violating decay mode, the presence of c-quarks can be exploited
by tagging c-jets. It is known that tagging c-jets is not easy because
of the low mass of c quark and the low decay length of the charmed mesons.
However, development of a strategy to tag the c-jets, even with 
a modest efﬁciency will be helpful to suppress the SM 
backgrounds by an enormous amount and hence needs to be 
pursued.

The paper is organized as follows.
In section \ref{sec2} 
a brief description of the parameter space
of interest is discussed. 
while in
sections \ref{sec4} and \ref{sec5} we discuss our collider strategy
 and results respectively along with a
discussion on the stop-coannihilation
 dark matter scenario.
  Finally  we conclude
in section \ref{sec6} .

\section{Parameter Space}
\label{sec2}
\subsection{CMSSM and PMSSM} 
We simulate the signal (Eq.\ref{eq1}) for the parameter space of  our interest in the context of both \newline CMSSM 
and the phenomenological
MSSM (pMSSM) with 19 parameters \cite{Djouadi:1998di}.
The CMSSM has been the most popular model of SUSY
 breaking in the context of collider phenomenology
and experimental searches over the last two decades.
 This is primarily
driven by the economy of the model which requires 
4 parameters and a sign, as compared to the 
cornucopia of over 100 parameters in the
 MSSM. These parameters defined at the GUT scale,  include
the universal scalar mass($\rm m_0$),
the universal fermion mass($m_{1/2}$) 
and the universal trilinear coupling $\rm A_{0}$,
along with $\rm tan\beta$, the ratio of 
vacuum expectation values(VEV) of the two Higgs
doublets and the sign of
 $\rm \mu$, the Higgsino mass parameter, 
at the weak scale.
The sparticle spectrum at the electroweak scale
is obtained by renormalization group running
from the GUT scale to the electroweak scale.

It is a well known fact the Higgs mass receives a substantial quantum 
correction resulting an enhancement of its mass from its tree level values
which is bounded by mass of Z boson, viz,
$\rm M_{H}^{2}\le M^{2}_{Z}cos^{2}{2\beta} $.
In order to accommodate the Higgs mass of 125 GeV in the CMSSM framework
one necessarily requires  stops 
in the multi-TeV regime or lighter stops with maximal mixing scenarios. Loop corrections 
can increase the tree level Higgs mass up to $\rm \sim$
140 GeV, due to
stop-top loops and a 
large value of the triline ar coupling 
$\rm A_{t}$\cite{Baer:2011ab,Akula:2011aa}. 
However it has also 
been noted that such large trilinear coupling 
introduces a significant amount of fine tuning in the theory.
On the other hand a large $\rm A_{0}$ results in a large splitting
in the stop mass matrix. This means that
lighter stops are accessible at LHC energies even 
in CMSSM \cite{Akula:2011aa}.
Hence it is worth investigating, in the CMSSM, the available parameter space,
which provides $\rm m_{H}\simeq 125~ GeV$ and where we have imposed 
various experimental and 
theoretical constraints as described below.
 
\subsection{Constraining CMSSM}
\label{sec:constraints}

With this goal, we perform a numerical scan of the relevant part of the
 CMSSM parameter space, varying the range of parameters such as,
\begin{equation}
 m_0 : [0-3~{\rm~TeV}],m_{1/2} : [0-1 {\rm~ TeV}], A_{0}: [-2-10~\rm TeV],  
\end{equation}
and setting the top quark mass to be 172.9 GeV.
We generate $5 \times 10^5$ random parameter points for a fixed value 
of $\tan\beta$. We fix the sign of the Higgsino mass parameter $\mu$ to 
be positive. We use the SusPect\cite{Djouadi:2002ze} spectrum generator 
to generate the masses of the supersymmetric particles for a fixed set 
of input parameters along with SuperIso \cite{Mahmoudi:2007vz}
for the calculations of the branching ratios of rare B-meson decays.

We refer to a point in the CMSSM parameter space to be allowed if it
survives the following constraints:
 
(i) The lightest Higgs boson mass $M_{H}$ falls in the window
$122.5~{\rm GeV} < M_{H} < 129.5~{\rm GeV}$.
Note that, owing to  the small difference in the central values given by 
the ATLAS and CMS collaborations \cite{:2012gu,:2012gk}, in order to 
be conservative, we used the number $126 \pm 2$ GeV as the experimental 
value. In addition, we added a 1.5 GeV theoretical uncertainty following\cite{Arbey:2012bp}.

(ii) The branching ratio (${\cal B}$) of the radiative decay $B \to X_s\gamma$
satisfies the following 95\% C.L. bound \cite{Asner:2010qj},
$$
2.6\times 10^{-4} < {\cal B}(B \to X_s\gamma) \ \ \ < 4.5\times
10^{-4} \ .
$$      

The HFAG Average and the SM predictions for this branching ratio are given by
$[3.55 \pm 0.24 \pm 0.09] \times 10^{-4}$ \cite{Amhis:2012bh} and $[3.15 \pm 0.23] \times  10^{-4}$\cite{Buras:2013ooa} 
respectively. We obtained the above 95 \% C.L. bound after 
including the intrinsic MSSM uncertainty of $0.15 \times 10^{-4}$, following the prescription given in \cite{Ellis:2007fu}.

(iii) The branching ratio of $B_s \to \mu^+\mu^-$ 
\cite{Aaij:2013aka,Chatrchyan:2013bka} 
lies in the 95 \% C.L. allowed range
$$
1.3 \times 10^{-9} < {\cal B}(B_s \to \mu^+\mu^-) < 4.5 \times
10^{-9}  \ .
$$
In order to get the above range we used the CMS+LHCb average and SM predictions to be $[2.9 \pm 0.7] \times 10^{-9}$ 
\cite{Aaij:2013aka,Chatrchyan:2013bka}
and $[3.56 \pm 0.18] \times 10^{-9}$ 
\cite{Buras:2013ooa} respectively.

Figures \ref{10-rn}  \& \ref{30-rn} show the results of our numerical study.

If a point is allowed by all of the three constraints above then the point is plotted in magenta (dark grey). A point is green (pale grey) if it satisfies only the constraint (i) but none of the other  two. Note that all the points (except the red (black) points in the 
$m_0$--$m_{1/2}$ plane) have been checked to satisfy the requirements of  electroweak symmetry breaking, electric and color neutral LSP, the LEP lower 
bounds on the masses and other theoretical consistencies e.g., absence of 
tachyonic states and so on. It is worth mentioning here that the impact of 
the measured higgs mass on the parameter space is rather strong; the region 
with low values of $m_0$ and $m_{1/2}$ is completely ruled out by this single 
measurement (i). The absence of green points for $\tan\beta$=10 shows that the 
bounds from (ii) and (iii) above are not strong enough to rule out any point 
which is not already disallowed by the Higgs mass. This situation changes gradually 
as we go towards larger $\tan\beta$ values as can be seen from Fig.~\ref{30-rn}. 
In Fig.~\ref{30-rn} there are many (green) points which are 
allowed by the Higgs mass but disfavored by the flavor physics data, in particular (iii). 
This happens because of the strong dependence of the branching ratios of 
$B \to X_s\gamma$ and $B_s \to \mu^+\mu^-$ on $\tan\beta$. 
In fact, the dominant Higgs contribution to these branching ratios are proportional 
to $(\tan\beta)^2$ (for ${\cal B}(B \to X_s\gamma)$) and $(\tan\beta)^6$ 
(for ${\cal B}(B_s \to \mu^+\mu^-)$).  This $(\tan\beta)^6$ dependence removes all the 
points with $m_0 < 3000$ GeV and $m_{1/2} < 1000$ GeV for $\tan\beta$ = 50, hence we 
do not show them here. 
We would also like to point out in passing that the dominant Higgs mediated diagrams 
mentioned above are also proportional to the stop tri-linear coupling $A_t$. Thus the 
constraints from (ii) and (iii) can be relaxed for small values of $A_t$. On the other 
hand, the measured value of the Higgs mass prefers a large value of $A_t$. Hence, there 
is a complementarity between the bounds coming from the Higgs mass and those coming 
from the flavor physics data.

In both Fig.~\ref{10-rn} and Fig.~\ref{30-rn} we also show (in the second row) the 
allowed values of $m_{\tilde{t}_1}$ for a wide range of $m_{1/2}$ and $A_0$ values. 
In spite of the constraints discussed above, it is quite clear from these figures that
even in the CMSSM, there exists regions in the parameter space where a light stop 
below a TeV mass scale is allowed. 
%

\subsection{Stop decay and benchmarks}

While it is certainly interesting to find a large 
region of parameter space pertaining to lighter 
stops and  allowed by the Higgs mass constraint,
the specific mass relations in CMSSM ties our 
hands to a large extent. In the context of natural
SUSY it is enough to consider only the third 
generation squarks (stops and sbottoms), the third
generation trilinear couplings along with
charginos and neutralinos. The rest of the spectrum 
is mostly unimportant and can be decoupled 
from this set. 
This rather simplified approach, in the framework 
of the phenomenological MSSM (pMSSM) brings out the relevant 
physics with the minimal number of input parameters.

The parameter space of our interest is guided by the region where
the flavor violating decay 
$\rm \widetilde{t}_{1}\to c \chi_{1}^{0}$,
 is kinematically dominant
for  relatively small mass differences 
 corresponding to Eq. \ref{eq:dm}.
 The decay width
is given by \cite{PhysRevD.36.724,Muhlleitner:2011ww},
\begin{equation} 
 \rm
\Gamma = \frac{1}{2}\alpha |\epsilon|^{2}m_{\tilde{t}_1}
\Bigg[1 - \frac{m_{\chi_{1}^{0}}^{2}}{m_{\tilde{t}_{1}}^{2}} \Bigg]^{2} ,
\end{equation}
where the loop factor $\rm \epsilon$ is
directly proportional to 
$\rm A_{b}^{2}$ and $\rm tan^{2}\beta$
with $\alpha$ being the strong coupling constant \footnote{For an exact one-loop calculation of this decay width see ~\cite{Muhlleitner:2011ww}.}.

A competing decay mode to the two body is the four
 body decay($\tilde{t}_{1}\to b\chi_{1}^{0}ff'$)\cite{PhysRevD.36.724,Djouadi:2001dx,Muhlleitner:2011ww},
dominantly via an off shell chargino into fermions. The two body decay 
mode, which is quadratically dependent on tan$\rm \beta$
and $\rm A_{b}$  dominates over the four body 
decay for moderate to high tan$\beta$.
 In the large $\rm tan\beta$
scenario, the two body decay mode  dominates over the four 
body decay mode, for similar mass differences as compared
to lower $\rm \tan\beta$.

\begin{figure}[h!]
\begin{tabular}{cc}
\includegraphics[keepaspectratio=true,scale=0.13]{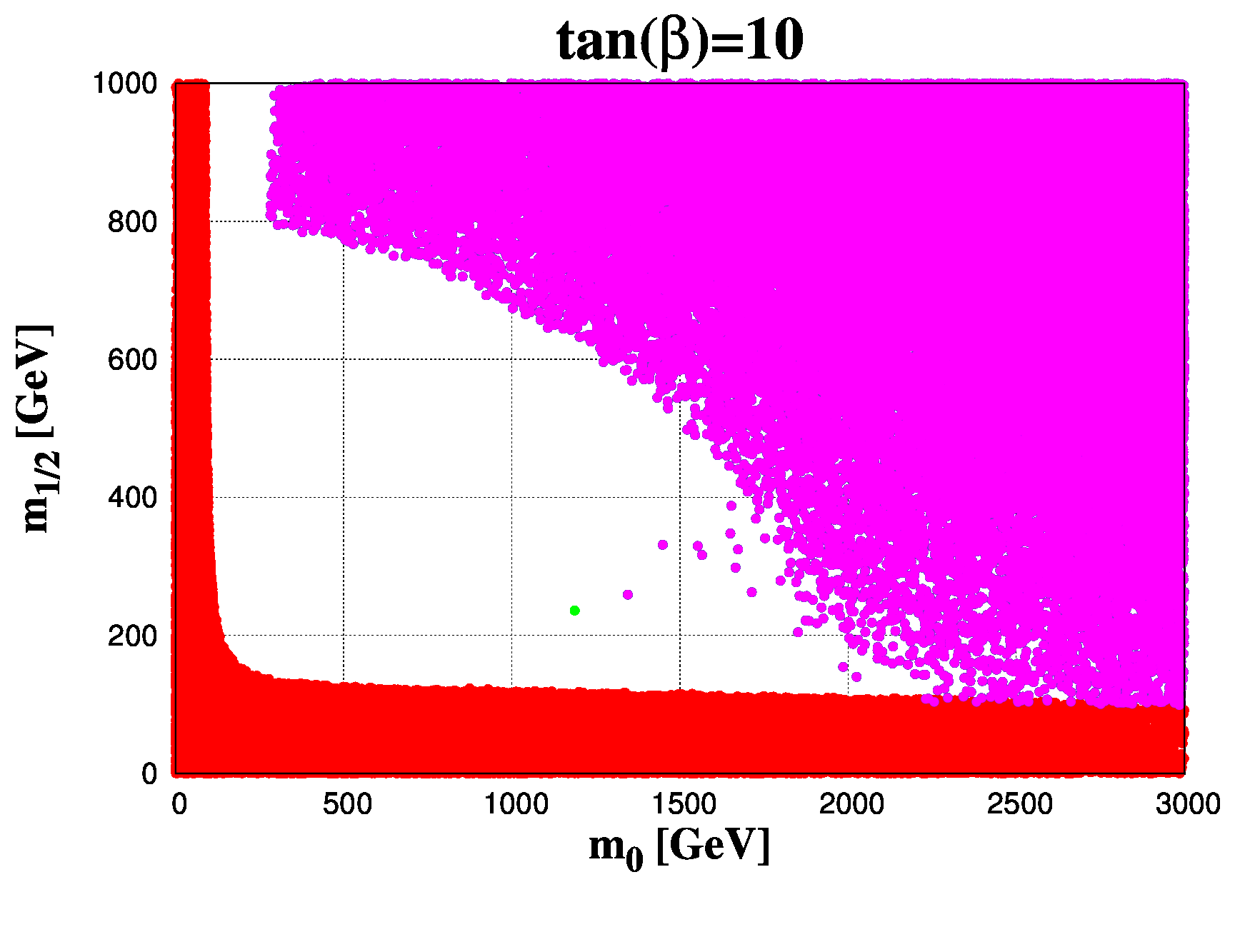}&
\includegraphics[keepaspectratio=true,scale=0.13]{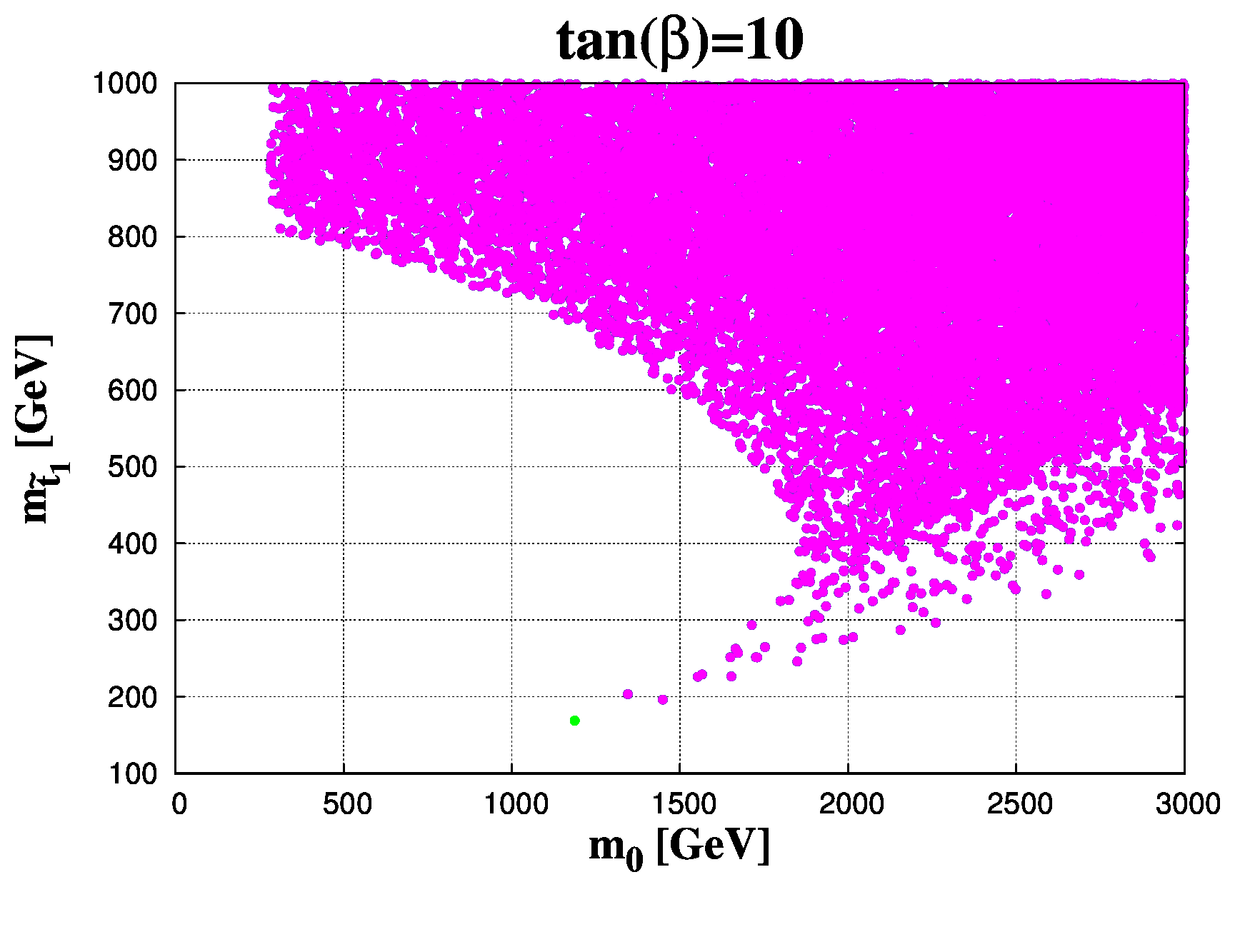} \\
\includegraphics[keepaspectratio=true,scale=0.13]{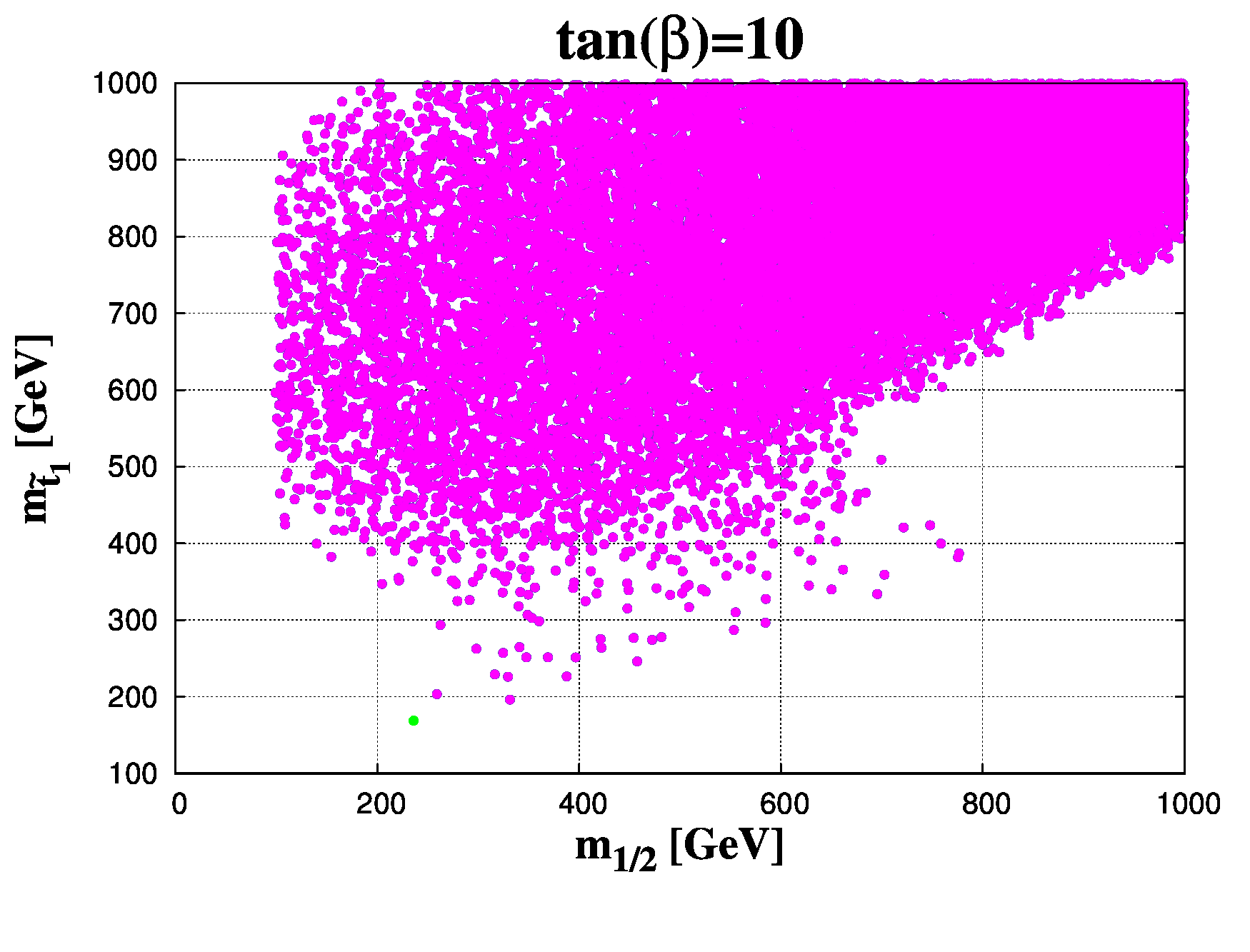}&
\includegraphics[keepaspectratio=true,scale=0.13]{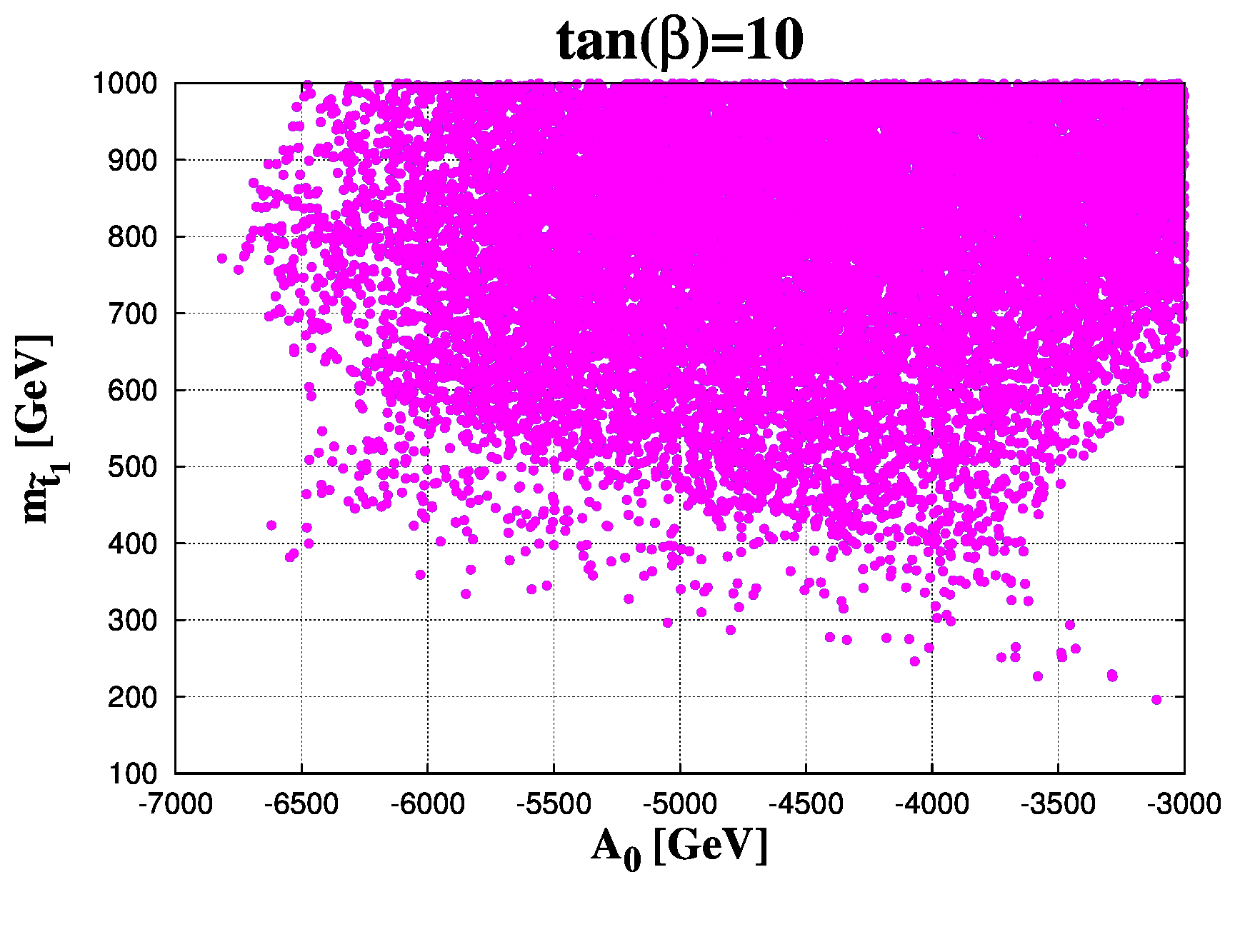}
\end{tabular}
\caption[]{The allowed parameter points in the $m_0 - m_{1/2}$ (top left panel), 
$m_0 - m_{\tilde{t}_{1}}$ (top right panel), 
$m_{1/2} - m_{\tilde{t}_{1}}$ (middle left panel), $A_{0} - m_{\tilde{t}_{1}}$ 
(middle right panel), in the CMSSM scenario for $\tan\beta=10$.  The red (black) region in the top left panel corresponds to points that are theoretically inconsistent.  All the three  constraints mentioned in the text are satisfied for the magenta (dark grey) points, whereas for the green (pale grey) points only the Higgs mass constraint is satisfied but the B-physics constraints are not. 
\label{10-rn}}
\end{figure}

\begin{figure}[h!]
\begin{tabular}{cc}
\includegraphics[keepaspectratio=true,scale=0.13]{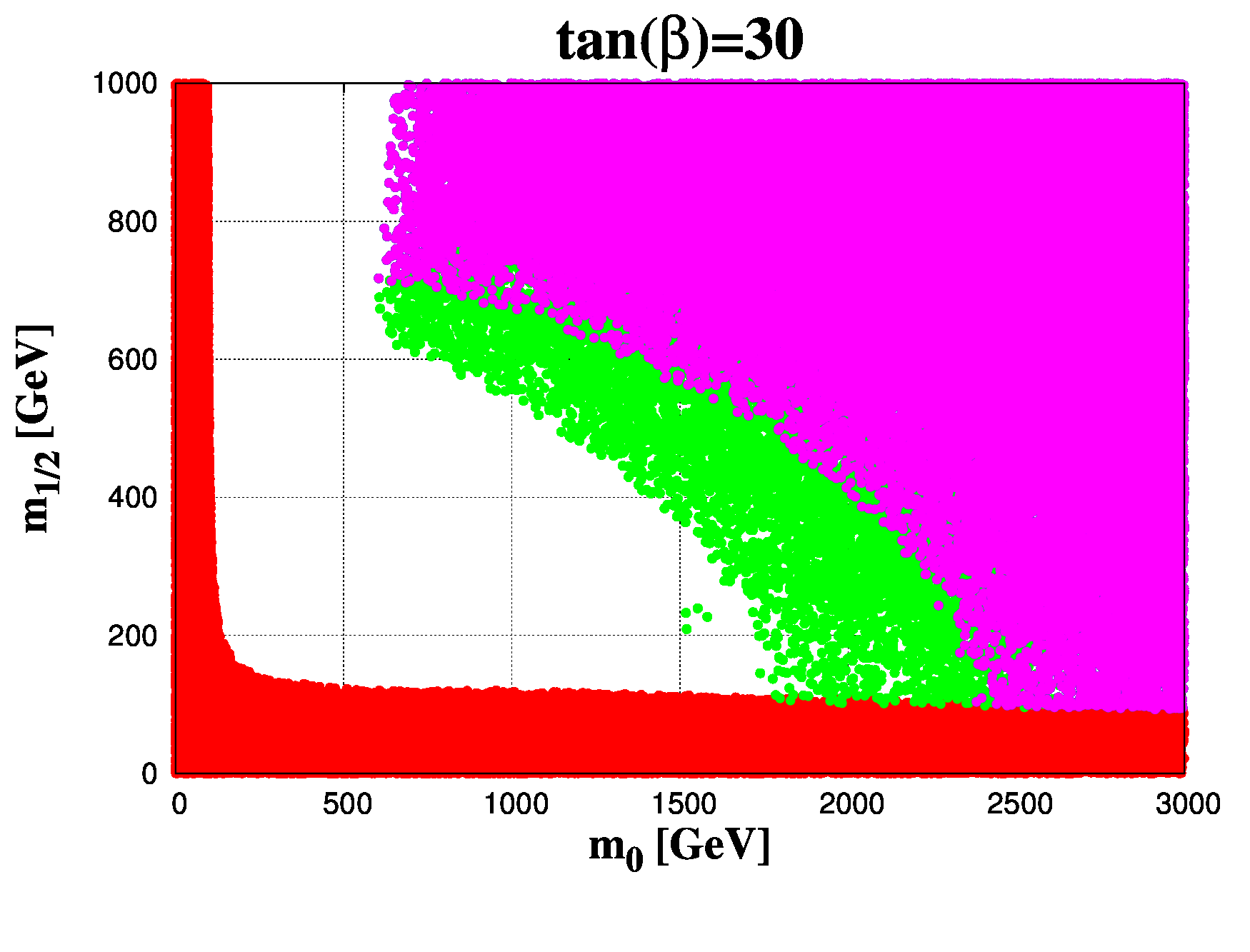}&
\includegraphics[keepaspectratio=true,scale=0.13]{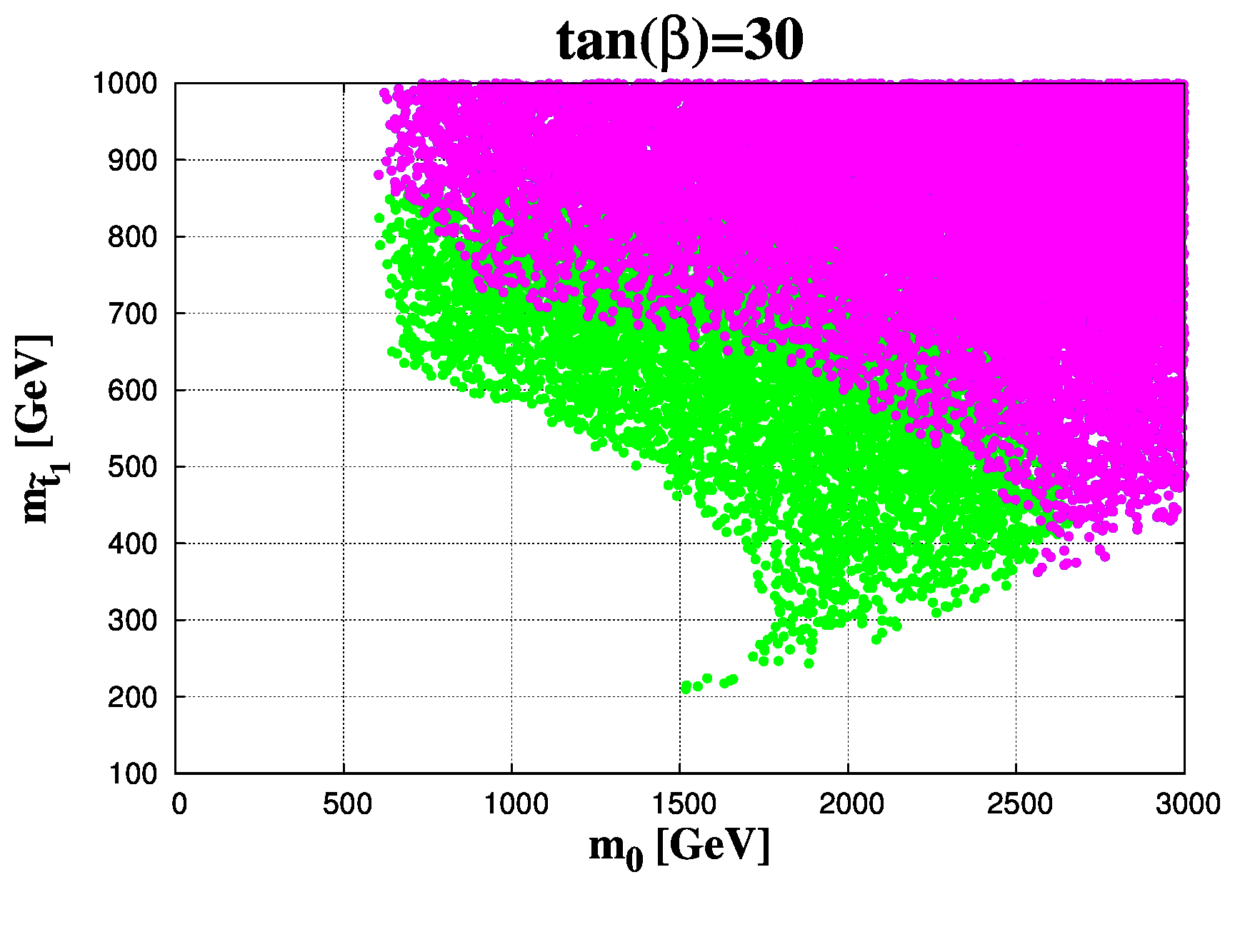} \\
\includegraphics[keepaspectratio=true,scale=0.13]{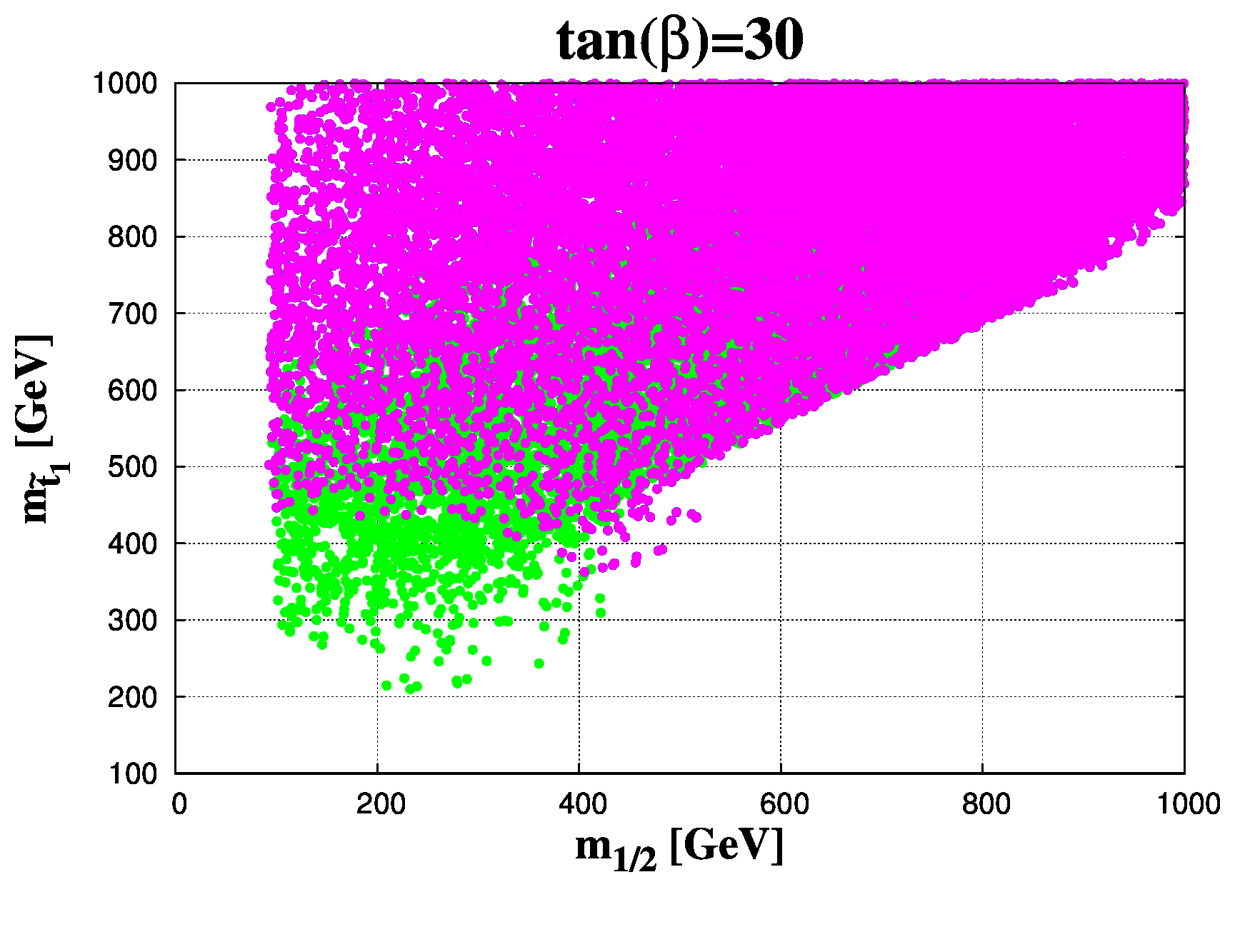}&
\includegraphics[keepaspectratio=true,scale=0.13]{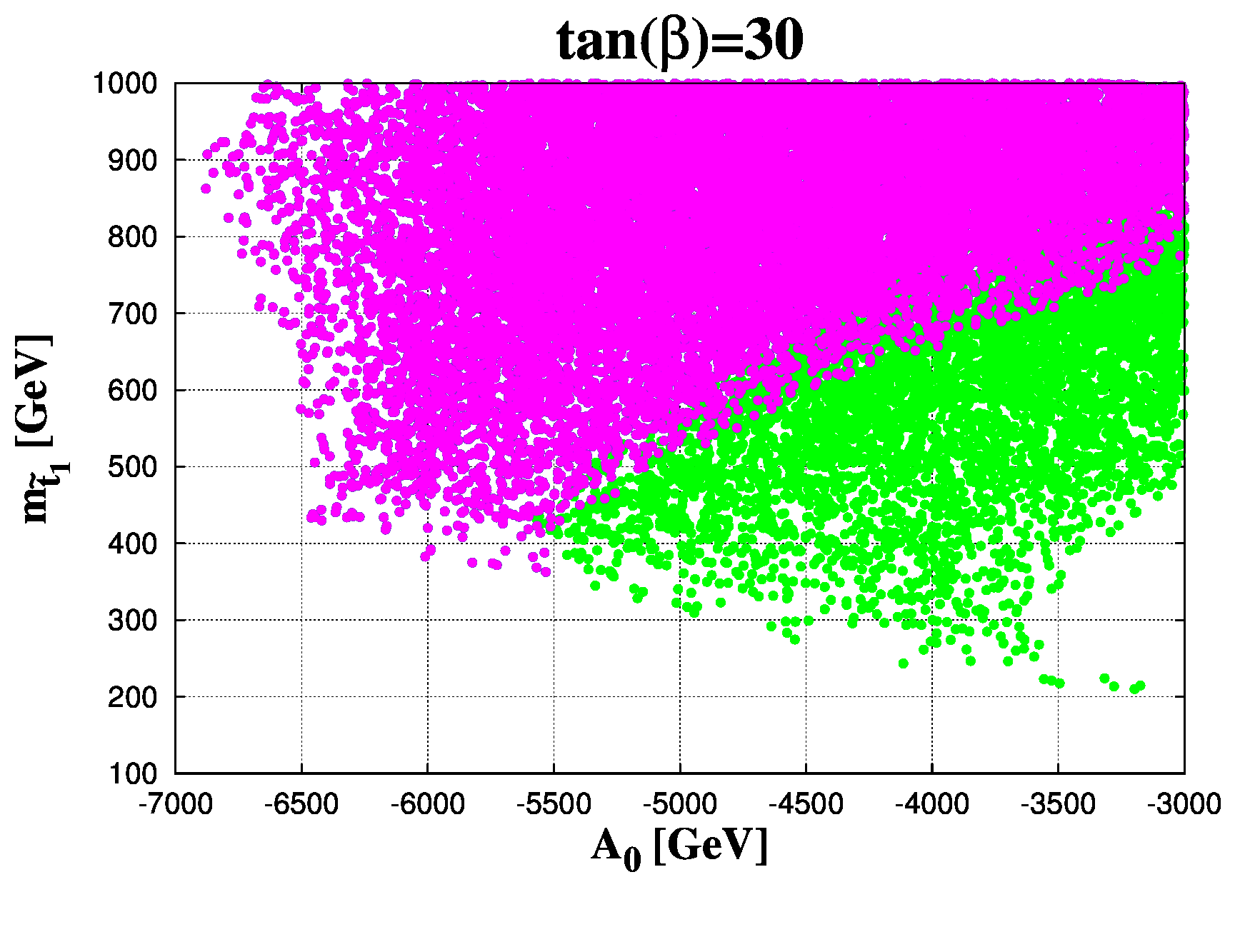}
\end{tabular}
\caption[]{The allowed parameter points in the $m_0 - m_{1/2}$ (top left panel), 
$m_0 - m_{\tilde{t}_{1}}$ (top right panel), 
$m_{1/2} - m_{\tilde{t}_{1}}$ (middle left panel), $A_{0} - m_{\tilde{t}_{1}}$ 
(middle right panel), in the CMSSM scenario for $\tan\beta=30$. 
 The colour code is the same as in Fig.~{\protect\ref{10-rn}}.
\label{30-rn}}
\end{figure}

\comment{
\textcolor{blue}{We choose  two benchmark points each in CMSSM and four points in the
19 parameter pMSSM models for our collider analysis.}}
 We choose  two benchmark points in the CMSSM and four points in the
19 parameter pMSSM for our collider analysis.
The representative points are shown in Tables \ref{tab1}
and \ref{tab2},
 indicating the relevant branching ratios and other sparticle masses.
All points in CMSSM (P1,P2) and three points 
in pMSSM(P3,P4,P5) are chosen to have 
$\rm tan\beta=10$.
We choose one additional benchmark point 
with tan$\beta$=30 in pMSSM(P6).

\begin{table}[h]
\begin{center}
\begin{tabular}{|c|c|c|c|c|c|c|c|c|c|c|}
\hline
 & $tan\beta$ &$m_{0}$ & $m_{1/2}$ & $A_{0}$& $m_{H}$ & $m_{\tilde{t}_{1}}$  & $m_{\chi_1^{0}}$ & $\Delta m$  &Br( $\tilde{t}_{1}\rightarrow c \chi_{1}^{0}$)  & Br($\tilde{t}_{1}\rightarrow bff'\chi_{1}^{0}$)   \\
  & & &  & & &   &  &  &$\%$ &  $\%$  \\
\hline
P1 & 10 & 1848 & 457.6 & -4069 & 126.0 & 241 & 198 & 43 & 74 & 25 \\
\hline
P2  & 10 &  2589 & 695 & -5849 & 126 & 331  &  306 & 25 &  97 & 2\\
\hline
 \end{tabular} 
\caption{ Masses  and branching ratios 
of some of the particles in the 
CMSSM scenario. 
All energy units in GeV. 
 $\rm sgn (\mu)$ is set to be 
 positive.}
\label{tab1}
\end{center}
\end{table}

\begin{table}[h]
\begin{tabular}{|c|c|c|c|c|c|c|c|c|c|c|c|c|c|}
\hline
&  $A_{t}$ &$tan\beta$ & $\mu$ & $M_{1}$ & $m_{H}$ & $m_{\widetilde{Q_3}}$  & $m_{\widetilde{t}_{R}}$
& $m_{\widetilde{b}_{R}}$  & $m_{\widetilde{t}_{1}}$  & 
$m_{\chi_1^{0}}$    &$\Delta m$ &
BR$(\widetilde{t}_{1}\to c\chi_{1}^{0})$ &  BR$(\tilde{t}_{1}\rightarrow bff'\chi_{1}^{0})$  \\
& & &  & & &   &  &&   &  &  &$\%$ &  $\%$  \\
\hline
P3 & -1900 & 10 & 800 & 280 & 123.0 & 380 & 1500 & 2000 & 355 & 285 & 70 & 98 & 2 \\
\hline
P4 & -2800  & 10 & 800 & 425  & 124.6 & 450 & 1800 &1800 & 458 & 432 & 26 & 96 & 3 \\
\hline
P5 & -2800  & 10 & 800 & 510 & 126.6 & 530 & 1800 & 1800 & 548 & 517 & 31 & 95 & 4  \\
\hline
P6 &-2800  & 30 & 800 & 425 & 128 & 500 & 1800 & 1800 & 520 & 432 & 88 & 98 & 2  \\
\hline 
\end{tabular} 
\caption{ Masses of some of the sparticles for the benchmark points in the
pMSSM scenario.
In all cases , the remaining parameters are as described in the text. 
All energy units are in GeV.
}
\label{tab2}
\end{table}

\comment{
\textcolor{blue}{For the parameter space corresponding to pMSSM in Table \ref{tab2},
we set the masses of the first two generation of squarks and all
slepton of all generations to
5 TeV and the gluino to 1.5 TeV as they are 
irrelevant for our study. The trilinear couplings
with the exception of $\rm A_{t}$ are set to zero. 
 All the benchmark points
have been checked against the previously mentioned constraints. }}
For the pMSSM benchmarks in Table \ref{tab2}, 
we set the first two generation of squarks and all
slepton generations to 5 TeV and the gluino to 1.5 TeV as they are 
irrelevant for our study. The parameter $M_2$ is set to 900 GeV.
 The trilinear couplings
with the exception of $\rm A_{t}$ are all set to zero. The pseudoscalar mass $m_A$ is set to 500 GeV. 
All the benchmark points have been checked against the constraints mentioned in Sec.~\ref{sec:constraints}.



\section{Signal and background}
\label{sec4}
 We simulate the collider signatures of
the stop pair production (Eq .\ref{eq1}) for the benchmark points as shown in  
Tables \ref{tab1} and \ref{tab2} in the dijet + $\PMET$ scenario
and the corresponding SM background at the LHC.
As pointed earlier, the final state objects
like jets and $\rm \PMET$ are expected to be 
soft  because of low $\Delta m$ making it difficult
to obtain a reasonable acceptance after cuts necessary 
to suppress the SM backgrounds. The signal cross section is rather 
small  at 8 TeV and  falls to
$\sim$ 50 fb for a stop mass of 500 GeV.
Most of the search strategies proposed have 
taken recourse to monojet + $\PMET$ or monophoton + $\PMET$ searches, where
a hard QCD jet is used along with a large $\PMET$ 
\cite{Drees:2012dd}.  It was demonstrated in 
\cite{Drees:2012dd} that with this strategy it is possible
to use this channel for stop discovery, for stop 
masses up to $\sim$ 300 GeV at 14 TeV LHC, 
with 100 $\rm fb^{-1}$ luminosity.

Corresponding to the signal
the principal SM backgrounds that can mimic the signal process are: 

\begin{itemize}
\item {\bf QCD }: The final state is swamped 
by the QCD dijet events, since the QCD 
cross section at hadron colliders is 
enormous. The $\PMET$ source 
in this case comes from semileptonic 
B-decays. There are non physics sources due to mismeasurement of jets,
detector noise
 which are 
out of the scope of this study.

\item {\bf Z($\rm \to \nu\bar{\nu}$) + jets} :
 This  makes up the irreducible
part of our background that looks exactly 
like the signal. Although the principal part of this
 background is Z + 2 jets, contribution from higher
jet multiplicities are not negligible
if some of the jets are not identified.

\item {\bf W($\rm \to l\bar{\nu}$) + jets } : 
This process contributes dominantly to the background 
when the lepton is not identified.
 Since the cross section for W + jets
is rather large this also contributes significantly
to the background.

\item {\bf $\rm t\bar{t}$} : This is primarily dominant 
when either the leptons from the W decay
 and/or some of the final
state jets are not identified leading to the same
configuration as the signal.

\item {\bf WW } : This process contributes to the background
when one W decays hadronically while the other leptonically.

\item {\bf WZ } : This again contributes substantially to the background
when W decays leptonically and Z decays hadronically 
with the lepton not being identified, or when 
Z decays to $\nu\bar{\nu}$ and W hadronically.

\item {\bf ZZ } :  This irreducible background mimics the 
signal in the situation Z($\to \nu\bar{\nu}$)Z($\to q\bar{q}$). 
\end{itemize}

We simulate the signal  and the
 background processes $t\bar{t}$, WW,WZ,ZZ using 
PYTHIA6 \cite{Sjostrand:2006za}. For the background processes W/Z+jets,
 parton level events are generated using 
ALPGEN \cite{Mangano:2002ea} and subsequently passed on to PYTHIA6
for showering and hadronization.  Jets are 
reconstructed  using FastJet \cite{Cacciari:2011ma} with 
anti-$\rm k_{T}$ \cite{Cacciari:2008gp} algorithm setting  a
size parameter R=0.5. Jets are selected 
with the  following criteria,

\begin{equation} 
\rm p_{T}^{j}\geq 30(60)~GeV~\forall ~8(13)~TeV,~|\eta|\leq~ 3.
\label{jetpt}
\end{equation}

 MLM matching \cite{Hoche:2006ph} is performed
 while showering parton level events using PYTHIA
for W/Z+ jets with a matching cone of 
$\Delta$R =0.7 and a jet $p_{T}$ threshold of 30 GeV within 
$\rm |\eta| \le$ 2.5.
We use CTEQ6L\cite{Lai:1999wy} as parton density
function(PDF)
from the LHAPDF\cite{Bourilkov:2006cj} package and
 set $Q^{2}=\hat{s}$.
Leptons are selected with,
\begin{equation}
 p_{T}^{l}\geq 10~GeV,~  |\eta|\le 2.5,
\label{leppt}
\end{equation}
which are used to veto events.

In order to suppress these backgrounds, in particular 
the large QCD di-jet background, we
use kinematic variable,
\begin{equation} 
\rm \alpha_{T}=p_{T}^{j_2}
/m^{jj}_{T},
\label{at}
\end{equation}
 where $\rm p_{T}^{j_2}$ is the transverse 
momentum of the second hardest jet and $\rm m_{T}^{jj}$
is the transverse mass of the two jet system 
\cite{Randall:2008rw}. It can be observed that for pure dijet
events, without any hard $\rm \PMET$ like QCD, the jets are
back to back in the transverse plane. Therefore the minimum
value of $m_{T}^{jj}$  in the limit
when jet masses can be ignored 
 turns out to be 2$p_{T}^{j}$ and thus 
the distribution of $\rm \alpha_T$ 
has a sharp end point at 0.5. 
 However for dijet events 
in association with a significant amount of $\PMET$,
as is the case for the signal in Eq. \ref{eq1}, the two jets 
are not back to back, leading to large values of $\alpha_T$.
We present the $\alpha_{T}$ distribution for signal and
 background in Fig. \ref{alphat} subject to
 the jet selection cuts ( Eq. \ref{jetpt}) 
 along with lepton veto. 
The signal process is for  P2 in 
Table 1 with $m_{\tilde{t}_1},m_{\chi_{1}^{0}}$
masses of 331 and 306 GeV respectively.
 The number of events generated to construct these plots 
correspond to the fourth column in Table \ref{tab3} (8 TeV).
 It can be seen in the figure, that as
predicted, the QCD process has a sharp fall at $\approx 0.5$
and therefore we impose a selection cut \cite{Randall:2008rw},
\begin{equation}
 \alpha_T >0.55.
\label{eqat}
\end{equation}
In addition we also 
use the kinematic variable $M_{T2}$\cite{Barr:2003rg,Lester:1999tx}
defined as, 

\begin{equation}
\rm M_{T2}(j_1,j_2,\PMET)= min~[max\{M_{T}(j_1,\chi),
M_{T}(j_2,\chi)\}],
\end{equation}  
 the minimization being performed 
over $\PMET^{1}+\PMET^{2}=\PMET$ where 
$\PMET^{1},\PMET^{2}$ are all possible partitions 
of invisible transverse momentum($\PMET$), which is due to the presence of LSP 
for signal and neutrinos for SM backgrounds. 
Here $\chi$ is the  invisible
particle whose mass ($M_{\chi}$) is an unknown parameter.
 The kinematic variable transverse mass($M_T$) between the jet and 
the accompanying missing particle is,
\begin{equation}
 M_T^{2}=M^{2}_j + M^{2}_{\chi} +2(E_{T}^{j}E_{T}^{\chi} - 
\vec{p}_{T}^{j}.\vec{p}_{T}^{\chi}) ,
\label{eq:mt}
\end{equation}
where  $\vec{p}_{T}^{j}$ is the transverse momentum vector of the jet
and $E_{T}^{j}$ the corresponding energy, while 
 $\vec{p}_{T}^{\chi}$ is the missing transverse 
momentum ($\PMET$) vector. 

 Since the maximum value of $M_T$ is restricted to the mass of the parent particle, 
$M_{T2}$ is also expected to be bounded by the respective parent particle mass.
Here $M_{T2}$ is calculated by setting $M_\chi$=0 without any loss of
generality~\cite{Barr:2009wu}. This assumption is clearly valid for SM processes
where the missing momentum is mainly due to neutrinos. Furthermore,
we found no  significant difference to the population of events near the end points 
for the signal with massive $\chi$ when we make this assumption.  However, there may be a difference
in acceptance for the two cases which we will consider as a systematic uncertainty 
in the acceptance efficiency.
In the SUSY processes where 
the parent particle($\rm \widetilde t_{1}$) is
heavier than SM particles, the tail in the $M_{T2}$ distribution is expected to extend up to
 a larger value.
  This variable thus provides
an excellent handle to suppress the remainder of the 
background rates.

In Figure \ref{mt2}, 
the signal and background distributions for 
$\rm M_{T2}$ are displayed for 8 (left)
and 13(right) TeV LHC energies. The distribution is
subject to a di-jet criteria along with 
a lepton veto with jet and lepton selection 
criteria(Eq.\ref{jetpt} and   \ref{leppt}) .
 The signal 
distribution displayed in the figure 
corresponds to the benchmark  P2 in Table \ref{tab1}.
 The number of events generated to construct these plots 
correspond to the fourth column in Tables \ref{tab3} (8 TeV) and 
\ref{tab4} (13 TeV).
 
We observe that for the signal process 
at 8 TeV the tail extends beyond the 
background processes and further reaches beyond
 the stop mass while for the background 
 the end point of the distributions correspond to lower values of $\rm M_{T2}$.
As the  missing energy in Z+jets comes from Z decaying
to a pair of neutrinos from the same  side of the configuration, 
the $\rm M_{T2}$ distribution is not 
expected to have an end point at the Z mass
which is reflected in this plot \cite{Barr:2009wu}.
 In case of $t \bar t$ background and the signal also, 
the distribution extends beyond the expected end point at the parent particle mass. The contribution of events in this region is dominantly due to the jets from hard radiations, in particular from final state radiations
\cite{Barr:2009wu}. This phenomena is more evident at 13 TeV
energy where objects are kinematically heavily boosted. 
By observing the distributions, we apply a cut at,
\begin{equation}
M_{T2} >{\rm 130~GeV(250~GeV)~for~8(13)~TeV}.
\label{eqmt2}
\end{equation}

This fact can be exploited to suppress backgrounds
when the signal is in unfavourable condition kinematically (eg. for small $\Delta m$).
 For example, for very small $\Delta m$, $\tilde t_1$ mostly 
decays invisibly, and hence $M_{T2}$ constructed 
out of this hard $\PMET$ and jets originating from
initial and final state radiation(ISR/FSR) which are un-correlated with 
the $\PMET$ has a much larger tail. For larger values of $\Delta m$,  
 the longer tail is not observed 
due to the fact that $\tilde t_1$ will have both  visible and the
invisible (LSPs) decay products for which the which momenta are correlated. 
In Fig.\ref{mt2dm} we present  the distribution for these two cases, with $\Delta m=$10 
and 140 GeV. Clearly $M_{T2}$ gives  some handle to recover 
sensitivity for the low $\Delta m$ scenario. It is to be noted here 
also that for the case $\Delta m=140 $ GeV, the decay
$\tilde{t}_1\to bW\chi_{1}^{0}$ may open up, 
 competing with the flavor violating decay mode.
Clearly the flavour violating decay mode with very low $\Delta m$ 
have some benefits due to ISR/FSR effects.
However, it is clear that these effects have a dependence on 
the models employed in the event generators.
 Therefore, in order to understand 
its effect in our signal sensitivity in a more precise manner one needs to 
do more detailed investigation which is postponed 
to a future work.

Finally, considering signal and background characteristics,
we use the following cuts to suppress the backgrounds.

\begin{figure}
\begin{center}
\includegraphics[height= 8 cm, width=9 cm]{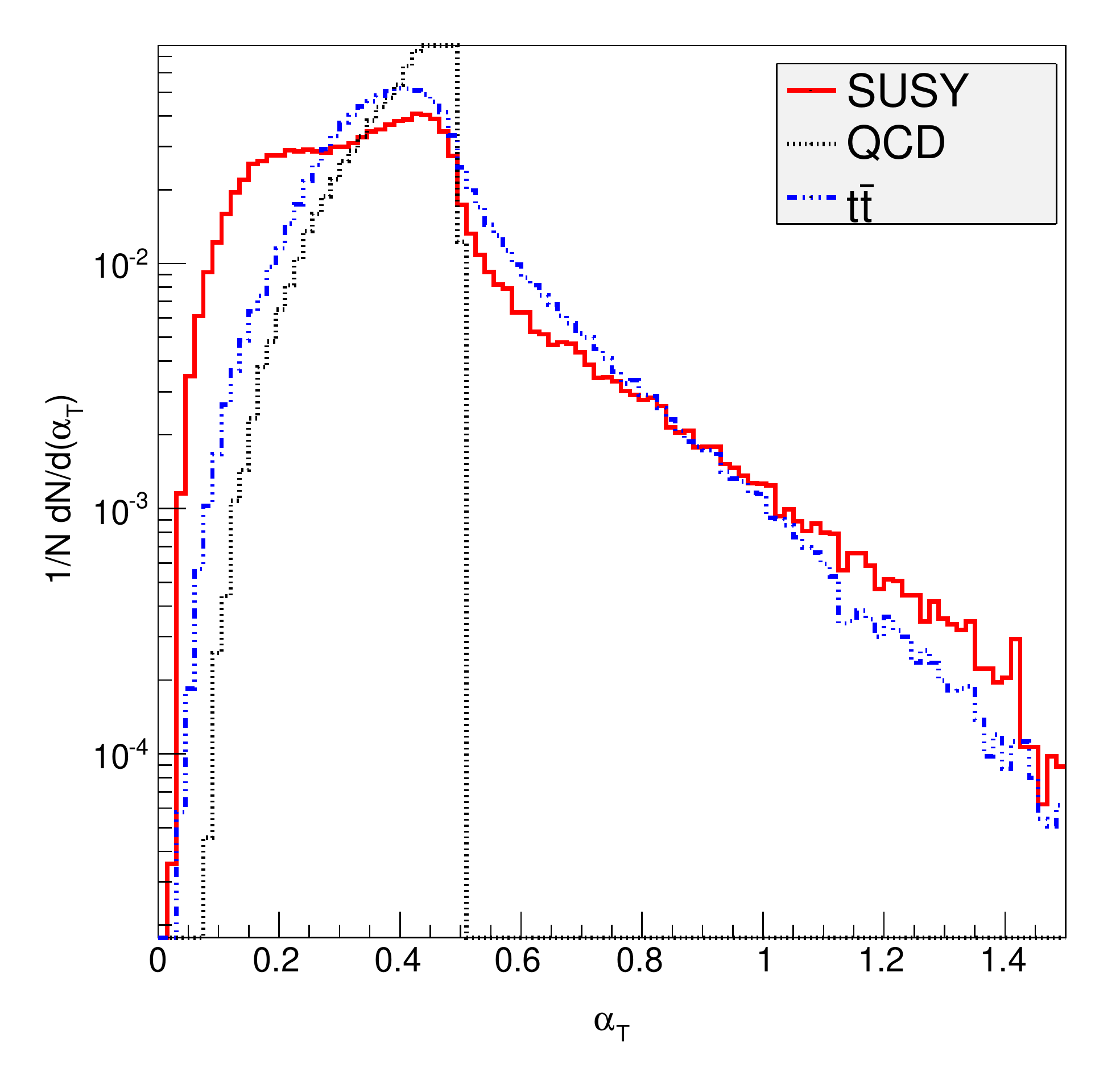}
\caption{$\alpha_{T}$ distribution for signal and background
at 8 TeV LHC energy. The signal corresponds  to  P2 from Table \ref{tab1}.
 The number of events generated correspond to the fourth column in Table \ref{tab3}.
}
\label{alphat}
\end{center}
\end{figure}

\begin{itemize}

\item {\bf Lepton Veto  (LV) :} In the signal process
 leptonic activity is absent. However 
background processes like $t\bar{t}$, W+jets,
WW,WZ contain a significant fraction of leptons
from W/Z decays accompanied by $\PMET$. The use of lepton veto thus helps to 
 suppress the backgrounds efficiently.
 Leptons are selected using cuts 
described in Eq. \ref{jetpt}.

\item {\bf 2 jets (2J) : } We select exclusively dijets with the 
jet $\rm p_{T}$ thresholds as described in Eq. \ref{leppt}. Note 
that 
after the lepton veto, the  $t\bar{t}$ background contribution is expected to be rich in  hadronic activity
and have more than 2 jets in the event. Hence a
strict imposition of the dijet criteria is expected
to reduce the background coming from the $t\bar{t}$, W/Z + jet processes.

\item {\bf b-jet veto (bJV) :} This veto is extremely efficient
in suppressing the top background.
The b-jet identification  is implemented by performing a matching
 of the jets with $b$ quarks assuming a matching cone
 $\Delta R(b,j) = 0.2$. 

\item {\bf $\rm \alpha_{T} \ge ~ 0.55$}  as discussed in Eq. \ref{eqat}.

\item {\bf  $\rm M_{T2} \ge~130~GeV $} (250 GeV for 13 TeV) as given by Eq. \ref{eqmt2}.

\begin{figure}
\includegraphics[scale=0.4]{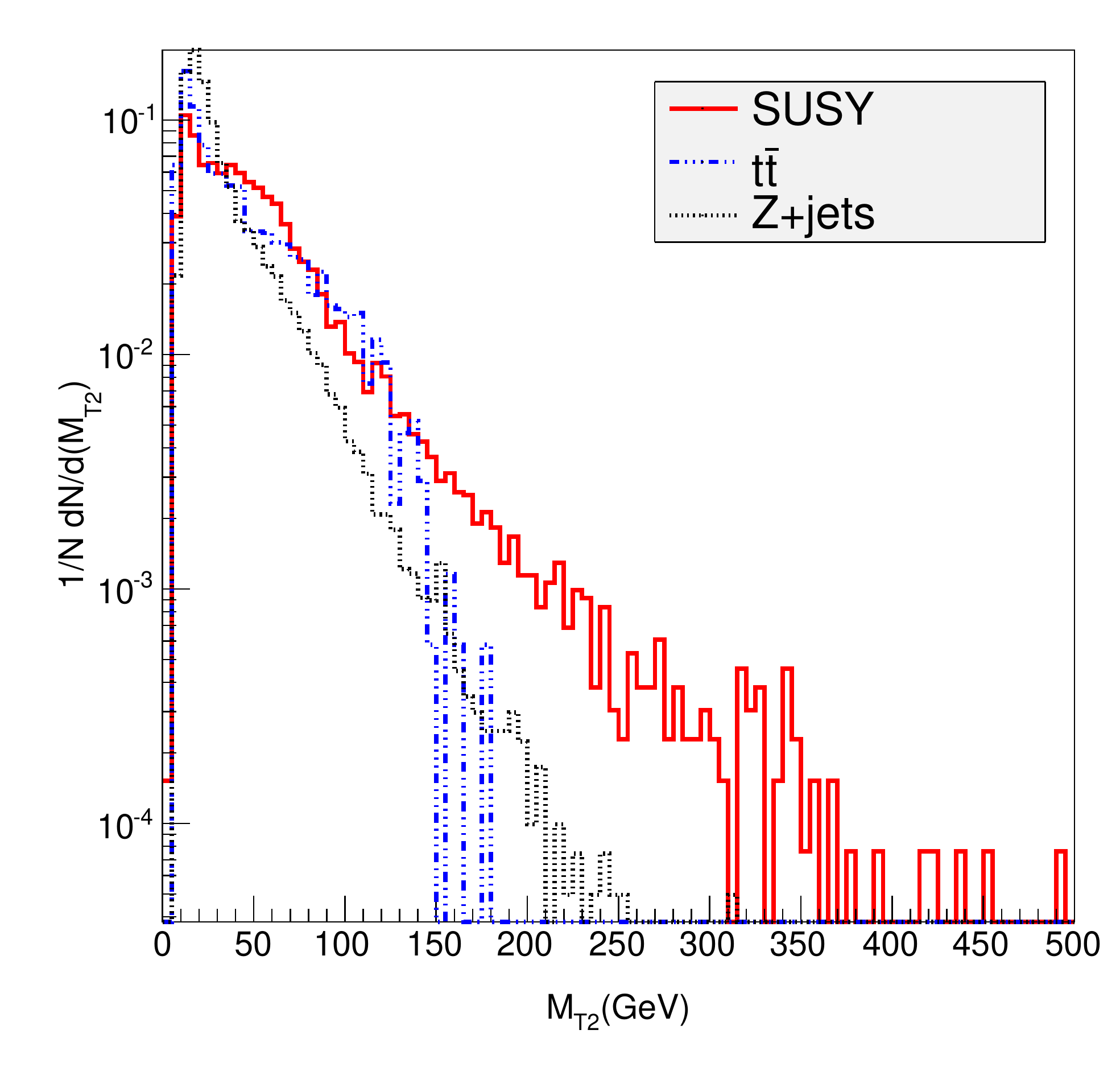}
\includegraphics[scale=0.4]{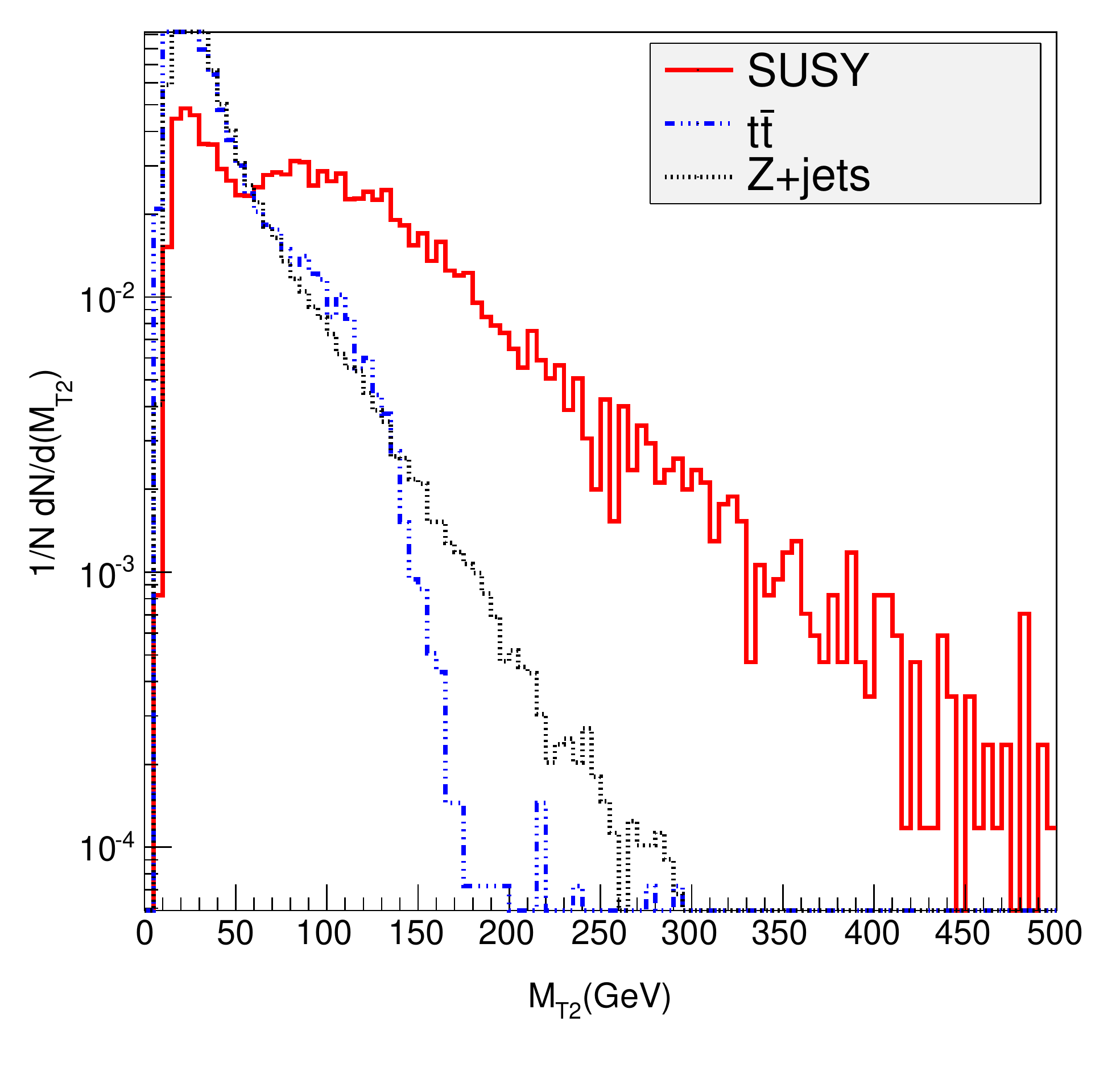}
\caption{$M_{T_2}$ distribution for signal and background for
8 TeV(left) and 13 TeV(right). The signal corresponds to 
 P2 in Table \ref{tab1}. 
 The number of events generated correspond to the fourth column in Tables \ref{tab3} (8 TeV) and 
\ref{tab4} (13 TeV).
}
\label{mt2}
\end{figure}

\item {\bf $\rm \PMET/H_{T}\le 0.9$ : } For signal
processes we expect this ratio to be less than 1,
while in background processes this is 
expected to be close to 1. The difference in
 azimuthal angle 
between the two jet system and missing energy
in signal and background being primarily 
responsible for this behaviour. We find that 
this selection is extremely effective 
in suppressing the $t\bar{t}$ background.

\end{itemize} 

At the end, we explore the possibility of  
an improvement  
by somehow tagging charm jets which are a 
part of the signal.  
 The identification of 
charm jet is quite challenging. Recently 
however, attempts have been made to 
measure W + c jets cross-section
where c-jets are identified\cite{CMS-PAS-SMP-12-002}.
Note that the channel $\tilde{t}_{1}\to c\chi_{1}^{0}$,
has been recently searched by ATLAS by 
trying to identify charm jets 
in the final state\cite{ATLAS-CONF-2013-068}.
 Although the method we employ 
is rather simplified, it still points 
out that charm like jet identification can prove
to be extremely effective in this case.
To identify charm jets we  match jets 
with charm quarks,
 using a matching
cone of R=0.2.
To find the presence of charm jets 
one can further check the
presence of D-meson among the jet constituents,
which we postpone to a future work. 

\begin{table}
\begin{tabular}{|c|c|c|c|c|c|c|c|c|}
\hline
Proc  & $m_{\tilde{t}_1},m_{\chi_{1}^{0}}$ & C.S & N  & LJV+2J+bJV   & $\alpha_{T}$  
& $M_{T2} \ge$  &  $\frac{\PMET}{H_{T}}$  & C.S  \\
 & GeV & &  &  & $\ge0.55$  & 130  & $\le0.9$    & (c-like)\\
\hline
P1 & 241,198 & 6330  & 0.5M  & 1996.2 & 262.3 & 4.7  & 2.1  & 0.45   \\
P2 & 331,306 & 1060  &  0.5M & 227.4  & 28.1  & 2.3  & 0.9  & 0.1  \\
P3 & 355,285 & 700   & 0.5M & 262    & 46    & 0.6  & 0.25  & 0.12   \\
P4 & 458,432 & 150   & 0.5M & 33.1   & 4.2   & 0.3  & 0.14  & 0.01  \\
P5 & 548,517 & 45    & 0.5M & 11.5   & 1.42  & 0.1  & 0.04  & 0.007 \\
P6 & 520,432 & 63    & 0.5M & 23.5   & 4.1   & 0.085& 0.04  & 0.02  \\
\hline
$\rm t\bar{t}$-5-200 & & 85000 & 2M  & 6063.4 & 80.24 & 8.9 & 0.46  & 0.17 \\
 $\rm t\bar{t}$-200-500 & & 9500 & 0.5M & 13.9 & 3.9 & 1.5 & 0.38  & 1 \\
$\rm t\bar{t}$-500- $\infty$ & & 130 & 0.5M  & 2.1 & 0.005  & 0.003  & 0.003 & $<1$ \\
\hline
qcd-300-500 &  & 1.3 $\times 10^{6}$ & 3M   & 37512.7 & $<1$ & $<1$  & $<1$   & $<1$\\
\hline
WW & &35000 & 0.5M  & 7462.2 &  380.2 & 2.94 & 0.84  & 0.14 \\
WZ & & 13000 & 0.5M  & 2547.8 & 189.3 & 3.1 & 0.76  & 0.21   \\
ZZ & & 5400  & 0.5M & 1050.1 & 78.12 & 1.62 & 0.3  & 0.02  \\
Z($\to \nu\bar{\nu}$)+2jet & & $10^{5}$ & 320910 & 71215.2 & 6877.4 & 67.07  & 29.4 & $<1$ \\
Z($\to \nu\bar{\nu}$)+3 jet & & 16500 & 241202 & 5349.6 & 637.0 & 9.9 & 4.5  & 0.13  \\
Z($\to \nu\bar{\nu}$)+4 jet & & 4240  & 39203 & 361.4 & 41.3 & 0.75 & 0.6    & $<1$  \\
\hline
W($\to l\nu$)+2jet & & $5.8 \times 10^{5}$ & 565087  & 117335.6 & 10752.1 & 63.4 & 35.3  & 2.1 \\
W($\to l\nu$)+3jet & &$10^{5}$ & 364287  & 14100.0 & 1240.8 & 9.1 & 6.1 &$<1$ \\
W($\to l\nu$)+4jet  & & 16300 & 44238  & 785.6 & 68.9 & 1.5 & 1.1 & $<1$ \\
\hline
\end{tabular}
\caption{ The cross sections(fb) for signal and backgrounds after
each cuts for 8 TeV LHC energy. The last two columns present normalized cross section(fb)
 after all cuts without and with identification of c-like jets respectively. All energy units are in GeV.}
\label{tab3}
\end{table}

\begin{figure}[t]
\begin{center}
\includegraphics[scale=0.5]{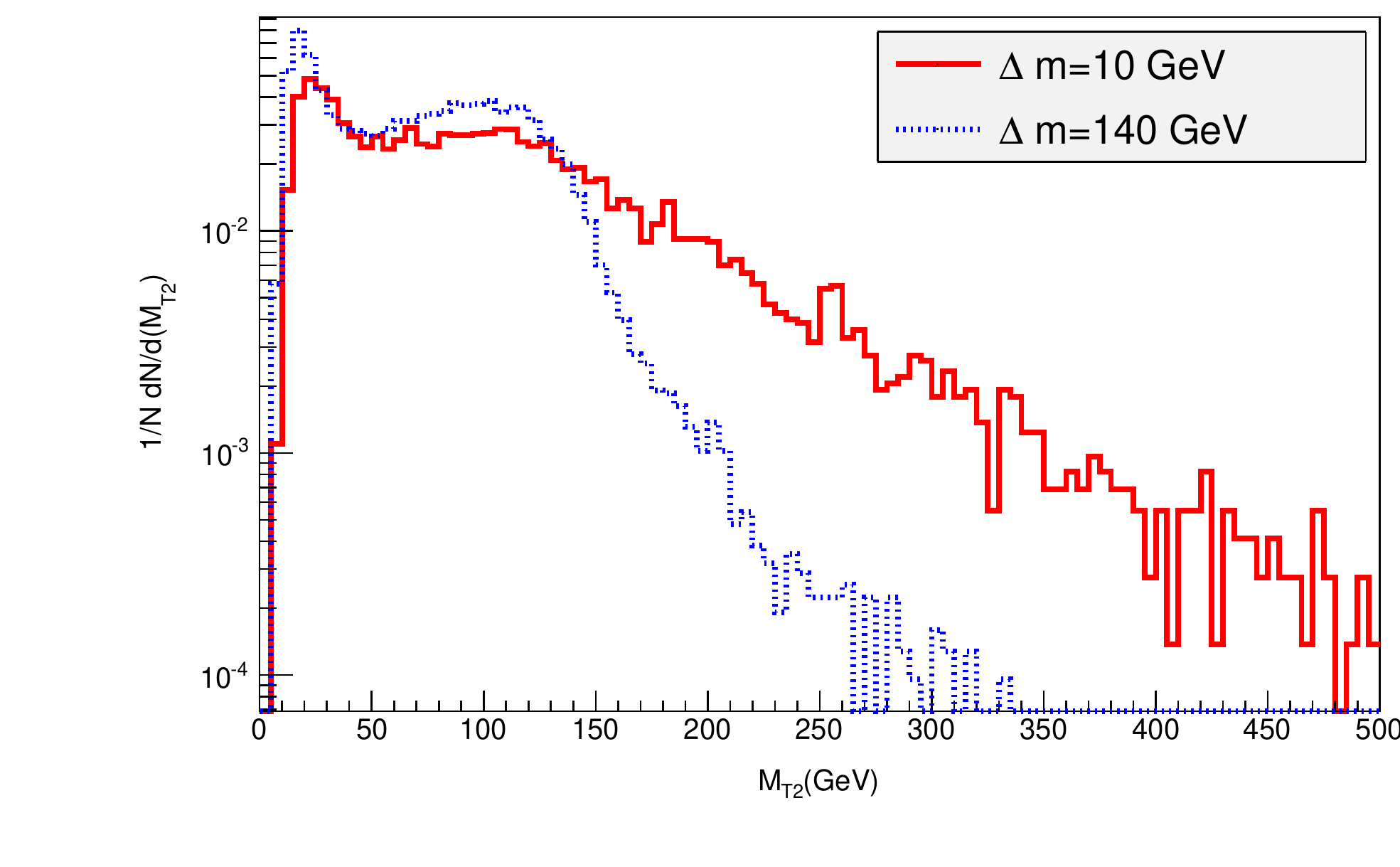}
\caption{$M_{T2}$ distribution for 
two values of $\Delta m$ of
10 GeV for solid(red) and 140 GeV for broken(blue) 
for a $m_{\tilde{t}_{1}}$ of 240 GeV at 13 TeV LHC energy.
 0.5 M events were generated to construct these plots.
}
\label{mt2dm}
\end{center}
\end{figure}


\section{Results}
\label{sec5}

Simulating the signal and background processes using the
selection cuts as described above, 
we present  results for 8 and 13 TeV LHC energies in
Tables \ref{tab3} and \ref{tab4} respectively.
  The first four columns
present the processes(Proc), the masses of
the lightest stop($m_{\tilde{t}_1}$) and LSP($m_{\chi_{1}^{0}}$),
 the cross section(C.S),  and the number of events generated (N) 
for each process
  respectively.
 We compute the 
next to leading order signal cross section 
using PROSPINO\cite{Beenakker:1996ed}.
 The subsequent columns display the cumulative effects of cuts, 
 as normalized cross sections (production cross section multiplied by cut 
efficiency) .
 In the penultimate column the cross 
sections after all cuts are presented.
 The top and the QCD backgrounds are 
simulated by slicing the entire phase space into $\rm \hat{p}_{T}$
bins, where $\hat{p_T}$ is the transverse momenta of the produced partons
in the partonic frame.
 In both Tables
we notice that the combined effects of lepton veto, b-jet 
veto and the dijet criteria (LV+2J+bJV) reduces the top
 background by an enormous amount ($\sim 95 \%$) while
 reducing the signal process
by about a third. 
 As pointed out earlier, the $\rm \alpha_{T}$ cut
successfully isolates the entire QCD background as expected.
   For the sake 
of simplicity we have quoted numbers corresponding to only
one $\rm \hat{p}_{T}$
bin for QCD.
 The rest of the backgrounds, particularly from top
and the WW/WZ/ZZ 
are also suppressed by a significant amount,  
costing a significant fraction of signal cross section as well. 
The $\rm M_{T2}$ cut,  as pointed out, removes a 
substantial fraction of $\rm t\bar{t}$,
WW/WZ/ZZ as well as W/Z+jets processes.
Clearly the $M_{T2}$ cut plays an important role
in isolating backgrounds efficiently.
 Finally the cut
$\PMET/H_{T}$  suppresses the WW/WZ/ZZ
backgrounds and brings it down to a negligible level. Even after a huge 
suppression of the  irreducible backgrounds W/Z+2,3 jets, 
the remaining fraction 
is non negligible because of the large production cross section.
Note that even after suppressing the SM backgrounds substantially,
since signal cross section is miniscule, 
the prospects of  discovering a signal at 8 TeV LHC is very limited.
In the last column cross sections are presented requiring 
that out of the two jets in the final state one is a c-like jet.
 As mentioned earlier, c-like jets are 
identified by naively  matching partonic c-quarks
and reconstructed jets. 
Clearly, it shows
that identification of c-like jets does help in reducing the
background to a large extent. This happens as the signal 
is likely to have a larger fraction of identified 
charm jets than the background.


\begin{table}
\begin{tabular}{|c|c|c|c|c|c|c|c|c|}
\hline
Proc &$m_{\tilde{t}_1},m_{\chi_{1}^{0}}$  & C.S & N  & LJV+2J+bJV 
& $\alpha_{T}$ & $M_{T2}\ge$ & $\frac{\PMET}{H_{T}}$ & C.S\\
 & GeV & &  &  & $\ge0.55$  & 250  & $\le0.9$   & (c-like)\\
\hline
P1 & 241,198 & 24100& 0.5M & 3113.2 & 311.6 & 8.1 & 4.4  & 0.45 \\
P2  & 331,306 & 4800 & 0.5M  & 440.2 & 82.0 & 5.8 & 3.2   & 0.18 \\
P3 & 355,285 &  3280 & 0.5M & 725.5 & 83.27 & 1.95 & 0.7  & 0.14 \\
P4  & 458,432 & 820 & 0.5M & 84.7 &  16.8 & 1.63  & 0.75  & 0.03\\
P5 & 548,517 & 290  & 0.5M & 34.2 & 6.2 & 0.5 & 0.29  & 0.023 \\
P6 & 520,432 & 400  & 0.5M  & 121 & 17.8 & 0.29 & 0.1  & 0.04  \\
\hline
$\rm t\bar{t}$-5-200 & & 291,000 & 4M  & 18690.2 & 2011.1 & 0.15 & 0.15  & $<1$ \\
$\rm t\bar{t}$-200-500 & & 39800 & 0.5M & 492.2 & 85.2 & $<1$ & $<1$   & $<1$    \\
$\rm t\bar{t}$-500-$ \infty$ & & 900 & 0.5M  & 12.4 & 0.13 &$<1$ & $<1$   & $<1$    \\
WW & & 69800 & 1M  & 11858.3 & 376.2 & 0.14 & $<1$   & $<1$ \\
WZ & & 26300 & 1M & 4144.9 & 231.5 & 0.5 & 0.15   & $<1$ \\
ZZ & & 10900 & 1M  & 1812.6 & 128.9 & 0.46 & 0.06 & $<1$  \\
\hline
Z($\to \nu\bar{\nu}$)+2jet & & 241,800 &2,605,885   & 34572.5 & 2643.4 & 12.4 & 1.3   & $<1$  \\
Z($\to \nu\bar{\nu}$)+3 jet & & 48000 & 418,467 & 14282.9 & 115.9 & 3.15 & 0.45   & $<1$  \\
Z($\to \nu\bar{\nu}$)+4 jet & & 11200 & 48473 & 3477.7  & 294.1 & 3.1 & 0.35  & $<1$   \\
W($\to l\nu$)+2jet & & 1,185,000 & 1,763,283  & 76017.1 & 3529.8 & 8.4 & $<1$   & $<1$  \\
W($\to l\nu$)+3jet & & 229,000 & 476,382 & 27524.6 & 1490.2 & 3.0 & $<1$  & $<1$ \\
W($\to l\nu$)+4jet & & 47200 & 103,178 & 6278.9 & 400.1 & 1.2 & $<1$   & $<1$  \\
\hline
\end{tabular}
\caption{Same as Table \ref{tab3} but for 13 TeV LHC.}
\label{tab4}
\end{table}

At 13 TeV the results improve significantly as can be 
seen from Table \ref{tab4}. The larger stop pair production cross section
 significantly helps in enhancing the event rates.
On the other hand an increased boost in the system  helps 
to effectively use the $M_{T2}$ variable
 by  applying a much larger cut value of 250 GeV to 
isolate the backgrounds. As can be observed from 
the right panel of Fig \ref{mt2}, a cut of 250 GeV 
is extremely effective in suppressing the irreducible Z+jets background.
We observe from the last column of Table \ref{tab4}
 that at 13 TeV energy, the signal and background cross
sections are comparable as compared to 8 TeV where the background
cross sections are dominant.

 Table \ref{tab5} summarizes cross sections of the signal and background
 after all cuts for 8 and 13 TeV LHC energy. 
The  $t\bar{t}$ cross section has been multiplied in the table by 
a k factor of 2 to take into account NLO effects
\cite{Kidonakis:2011tg}. Note that the k-factors for
 W/Z + jets processes are very close to 1 \cite{Berger:2010zx,Ita:2011wn}, 
and hence do not change our results significantly.
From Table \ref{tab5} we find that at 8 TeV the total background is 
70.6 fb 
in which the dominant contribution comes from W/Z + jets,
 while the signal cross section varies from 2.1 fb  for  P1 
to 0.04 fb in P6. 
Therefore for P1 with 20 $fb^{-1}$ luminosity
we obtain $S/\sqrt{B}=1.1$, while the significance drops substantially
with the increase of $m_{\tilde{t}_1}$. 
 At 13 TeV the total background cross section turns
out to be 2.6 fb while the signal cross section varies from
4.4 fb in P1 to 0.1 fb in P6. Thus for the points P1 and P2 
we obtain $S/\sqrt{B}=$~6.1 and 4.4 respectively 
for 5 $fb^{-1}$ which implies that P1 is discoverable
while an evidence of a signal can be obtained even with 
low luminosity options for P2. With 100 $fb^{-1}$ luminosity
we find that the benchmarks P3 and P4 have significance values of
4.7 and 4.3 respectively.

With this strategy we attempt to explore the sensitivity 
in $\Delta m $ for various values of $m_{\tilde{t}_1}$
at a given luminosity. We present our findings in 
Fig. \ref{eff} where 
the accessible region below
 the curves are shown in the 
$m_{\tilde{t}_1}-\Delta m$ plane for two luminosity
options 5 $\rm fb^{-1}$ and 100 $\rm fb^{-1}$ with $S\sqrt{B}\ge 5$. This plot is
presented with the assumption of $BR(\tilde{t}_1\to c\chi_{1}^{0})= 100 \%$, 
 holding the rest of the parameters fixed as described in section \ref{sec2}.
  The plot is obtained by a parameter space scan in the $m_{\tilde{t}_{1}}-\Delta {m}$ plane
with a grid spacing of 20 GeV on the $m_{\tilde{t}_{1}}$ (X)-axis and 5 GeV on the $\Delta {m}$ (Y)-axis.
 We find 
that even for low luminosity a light stop up
 to a mass of $\sim$ 350 GeV could be explored for $\Delta m=20 GeV$.
As mentioned earlier this search strategy is very sensitive to lower values of
$\Delta m$ and it is reflected in the figure. For a
luminosity of 100 $fb^{-1}$ we find that a light stop up to
a mass of 450 GeV can be probed for $\Delta m$ as low
as 35 GeV. 
The solid horizontal line demarcates the kinematic region 
$m_{\tilde{t}_1}< m_{t} +m_{\chi_{1}^{0}}$, 
over which the the decay $\tilde{t}_1\to t \chi_{1}^{0}$
opens up and dominates. 

\subsection{Implications for  dark matter}

In scenarios with  a small stop-neutralino mass splitting and a bino($\tilde{B}$) LSP,
stop-coannihilation  can play an important role in determining the relic dark matter abundance. 
This is the case especially for small $\Delta m$, i.e, where the decay
$\tilde t_1 \rightarrow c \tilde \chi_1^0$ is dominant.
Hence it is important to investigate the implications of our 
studies for probing the stop coannihilation scenario at the LHC.
\begin{table}
\begin{center}
\begin{tabular}{|c|c|c|c|c|c|c|c|}
\hline
  &  P1   & P2 & P3  & P4  & P5 & P6   & Total Bg\\
\hline
$m_{\tilde{t}_1},m_{\tilde{\chi_1^{0}}}$(GeV)& 241,198 & 331,306 & 355,285 & 458,432 & 548,517 & 520,432 &  \\
\hline
 8 TeV  & 2.1 & 0.9 & 0.25 & 0.14 &  0.04 & 0.04  & 70.6 \\
$S/\sqrt{B}$($20 fb ^{-1}$) & 1.1 & 0.5 & 0.13 & 0.07 & 0.02 & 0.02 &  \\
\hline
 13 TeV & 4.4 & 3.2 & 0.7  & 0.75 &  0.29 &0.1     & 2.6  \\
 $S/\sqrt{B}$(30 $\rm fb^{-1}$) & 15 & 11 & 2.4 & 2.5 & 1 & 0.3 & \\
\hline
\end{tabular}
\caption{
The signal and backgrounds cross sections(cs) for benchmark points.
 The signal significances($S/\sqrt{B}$) for different 
energies and luminosities are also shown.
}
\label{tab5}
\end{center}
\end{table}

The relic density crucially depends on the stop-neutralino mass difference as well as on other parameters that will be discussed in the next paragraph.
For the benchmarks considered, the value of the relic density as shown in Table \ref{tab:DM} is either above (P1,P3,P6) or below (P2,P4,P5) the central value for PLANCK,~$\Omega_{DM} h^2=0.1199$~\cite{Ade:2013zuv}.
However it is well known that the value of the relic density in coannihilation scenarios depends critically on the NLSP-LSP mass difference, hence   a small decrease (increase) in the stop-neutralino mass difference will lead to a large decrease (increase)  in  the value of the relic density. We have therefore searched for modified benchmarks for which  the relic density was in agreement with the central value of PLANCK. For this, we vary  only the mass of the lightest neutralino by changing $M_1$, while keeping all other parameters of each benchmark P1-P6 to their value at the EWSB scale.  The modified benchmarks, P1'-P6', with the corresponding stop and neutralino masses are listed in Table~\ref{tab:DM}.  The relic density is calculated using  $\rm micrOMEGAs3$ \cite{Belanger:2013oya}. Furthermore, their position in the $\Delta m - m_{\tilde{t}_1}$ plane is  displayed in Figure ~\ref{eff}.
We find that for the benchmark points that satisfy the PLANCK
constraint our search strategy works very well indeed. We achieve
reasonable sensitivity for stop masses below 400 GeV with an integrated luminosity  $\mathcal{L}~=~100~ {\rm fb}^{-1}$  at 13 TeV LHC as can be seen
in Fig.~\ref{eff}. In fact the  $5\sigma$ significance contours for  $\mathcal{L=~}5~ {\rm fb}^{-1}$ even covers the relevant $\Delta m$ for stop masses below 280 GeV.

A few comments are in order to ascertain the generality of this result since the relic density depends not only on the stop-neutralino mass difference but also on the nature of the neutralino LSP, the nature of $\tilde{t}_1$ (whether it is dominantly LH or RH), and on the value of $M_A$. First note that the mass splittings associated with the modified benchmarks of Table ~\ref{tab:DM} are typical of scenarios where the LSP is a bino, and these are precisely  the ones where stop coannihilation plays an important role in obtaining a low enough relic density. 
Second, the mass splitting  required to satisfy the PLANCK constraint - for a given stop mass - should depend  on whether the stop is LH (P3'-P6') or RH (P1'-P2'). The reason is the following: co-annihilation processes such $\tilde\chi_1 \tilde{t}_1 \rightarrow t g,th$ have a
larger cross-section for a RH stop than for a LH stop of the same mass since the coupling to the bino LSP is proportional to the
top hypercharge (which is larger for the RH top/stop), hence one would expect the required $\Delta m$ to be larger for a RH stop. Furthermore the QCD processes involving pairs of squarks $\tilde{t}_1\tilde{t}_1\rightarrow t t, \tilde{t}_1\tilde{b}_1\rightarrow tb ...$ which are more important for LH stops  involve
two Boltzmann suppression factors \footnote{The Boltzmann factor is $e^{-\Delta m}/T_f$ for each coannihilating particle and $T_f$ is the freeze-out temperature.}, therefore the mass splitting required is smaller. 
However since the Boltzmann factor varies rapidly with $\Delta m$, in the end there is only a few GeV differences between
the case of the RH and LH stop.  For example for benchmark P2', the required mass splitting would
be $\Delta m=29 {\rm GeV}$ for a dominantly LH stop instead of  $\Delta m=37 {\rm GeV}$  for a RH one. 
Finally, the pseudoscalar mass, $M_A$,  can also be a relevant parameter. For P1'
and P2' it is set by CMSSM boundary conditions and is rather high hence plays no role  in neutralino annihilation while for
 P3'-P6', it is set to 500 GeV. A higher value of $M_A$ would not affect
our collider search strategy and the relevant branching ratios of $\tilde t_1 \rightarrow c \chi_{1}^{0}$. However it
 would  imply smaller
mass differences than the ones listed in Table ~\ref{tab:DM}, and hence would easily be covered by our search strategy.
We can therefore safely conclude that the channel
investigated here can probe the stop-coannihilation scenario for stop masses up to at least 400 GeV.

\begin{table}
\begin{center}
\begin{tabular}{|c|c|c|c|c|c|c|}
\hline
  & P1   & P2 & P3  & P4  & P5 & P6  \\ 
  \hline
   $m_{\tilde t_1}, m_{\tilde\chi_1^0}$ (GeV)& 241,198   &331,306     &355,285    &458,432    &    548,517    & 520,432 \\
 $\Omega h^2$  & 0.17 & 0.04 & 1.9 & 0.04 &  0.06 & 0.59  \\
 \hline\hline
 
  & P1'   & P2' & P3'  & P4'  & P5' & P6'  \\ 
  \hline
  $m_{\tilde t_1}, m_{\tilde\chi_1^0}$ (GeV)& 241,205   &331,294     &355,315    &458,420    &    548,508    & 520,479 \\
 $\Omega h^2$  & 0.119 & 0.119 & 0.119 & 0.119 &  0.119 & 0.119  \\
 $\Delta m ~(GeV)$        & 36 & 37 & 40 & 38 & 40 & 41 \\
 \hline
\end{tabular}
\caption{Relic density for the benchmarks P1-P6 and the modified benchmarks P1'-P6'}
\label{tab:DM}
\end{center}
\end{table}

\section{Conclusion}
\label{sec6}

 In this study we perform a comprehensive 
analysis of the collider search prospects
of the flavor violating decay of the stop
quark, namely $\tilde{t}_{1}\to c\chi_{1}^{0}$.
Such a scenario is well motivated in the 
context of natural SUSY as well as 
from the dark matter perspective of 
stop co-annihilation. It had been 
earlier observed that this channel
is rather difficult to probe due 
to the low mass difference between 
the stop quark and the lightest neutralino.
The principle background to this 
channel arises from QCD, $t\bar{t}$ and
 Z($\nu\bar{\nu}$)/W($\to l\nu$)+jets 
final state.
We use the kinematic variables 
$\alpha_{T}$ and $M_{T2}$ to 
effectively suppress these
 at 8 and 13 TeV LHC.
 At 8 TeV, the level of
background is still high and we are 
limited by 
low stop pair production cross section. We find 
that our strategy is far more
effective at 13 TeV due to the increase in 
cross section and efficient use 
of the kinematic variables.  We observe that 
it is possible to discover light stop
quarks up to a mass of $\sim$ 450 GeV 
with 100 $fb^{-1}$ luminosity at 13 TeV
LHC energy for low values of $\rm \Delta m$
 and even for the case when the
$\tilde{t}_1$ and $\chi_{1}^{0}$
are almost degenerate.
We observe that for very low
$\Delta m$ case, the loss of acceptance because of soft visible
particles in the final states is compensated by ISR/FSR effects through 
$\rm M_{T2}$ selection.
 The 
result improves significantly 
when one attempts identifying 
charm jets both at 8 and 13 TeV LHC.
We also show the discoverable region 
in $m_{\tilde{t}_1}-\Delta m$ plane 
assuming the branching ration of $\tilde{t}_1\to c\chi_{1}^{0}$
to be 100$\%$.
This is an useful information in the context 
of DM via stop
co-annihilation. 
Our analysis shows that a good region of the parameter space
relevant for the stop - coannihilation scenario
can be probed at 13 TeV LHC energy.

\begin{figure}[t]
\begin{center}
\vspace{-2.5 cm}
\includegraphics[scale=0.6]{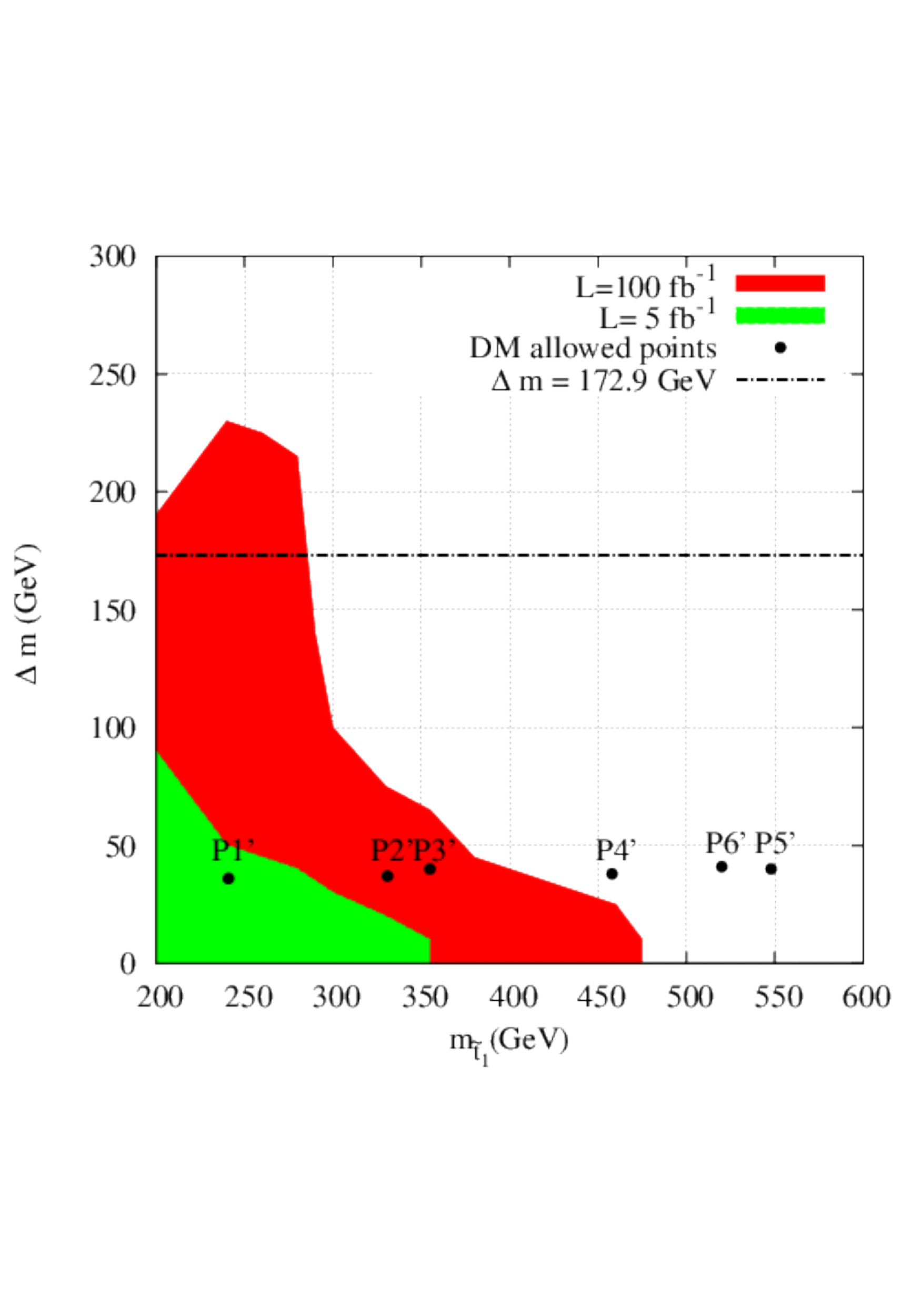}
\vspace{-3.0 cm}
\caption{The 5 $\sigma$ significance 
contours for $\rm \mathcal{L}=5~fb^{-1} $ \{green(light solid)\}, and 
for  $\mathcal{L}=100$~ $fb^{-1}$ \{red(dark solid)\} 
 luminosity assuming $\tilde{t}_{1}\to c\chi_{1}^{0}$
to be 100 $\%$ for 13 TeV LHC energy. 
The black (broken) line corresponds to $m_t=172.9~GeV$
and is the kinematic limit for
$\tilde{t}_1\to t + \chi_{1}^{0}$. The dark matter allowed points 
P1'-P6' corresponding to Table \ref{tab:DM} are denoted by the
black (solid) dots.}
\label{eff}
\end{center}
\end{figure}

\section{Acknowledgements}
The authors would like to thank the organizers
of WHEPP-XII held at Mahabaleswar,India in 2012, and the
BSM working group, where this project was initiated.
GB  thanks the Aspen Center for Physics for hospitality during the final stage of this work.
Partial funding by the French ANR~DMAstroLHC is gratefully acknowledged.
The research by DG leading to these results has received funding from the European Research Council under the European Union’s Seventh Framework Programme (FP/2007-2013) / ERC Grant Agreement n.279972.
 RMG
wishes to acknowledge the Department of Science and Technology
 of India, for financial support under the
J.C. Bose Fellowship scheme under grant no. SR/S2/JCB-64/2
007.

\end{large}

\bibliography{stop-journal.bib}{}
\bibliographystyle{utphys.bst}

\end{document}